\documentclass[]{article}

\usepackage{times}
\usepackage{geometry}
\usepackage[utf8]{inputenc}
\usepackage{enumitem,amssymb}
\usepackage{ragged2e}
\usepackage[citecolor=blue]{hyperref}
\usepackage{graphicx}
\usepackage{natbib}
\usepackage[T1]{fontenc}
\usepackage[polish,english]{babel}
\usepackage[utf8]{inputenc}
\usepackage{aas_macros}
\setcitestyle{aysep={}}   

\newlist{thematic}{itemize}{8}
\setlist[thematic]{label=$\square$}
\usepackage{pifont}
\newcommand{\rev}{\textcolor{black}}

\newcommand{\orcid}[1]{\href{https://orcid.org/#1}{\includegraphics[width=10pt]{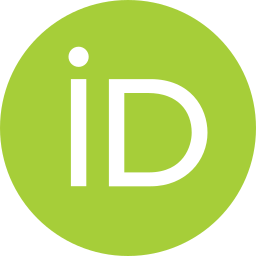}}}

\hypersetup{
    colorlinks=true,
    linkcolor=blue,
    filecolor=magenta,      
    urlcolor=cyan,
    pdfpagemode=FullScreen,
}

\begin{document}

\raggedright
\LARGE
Bringing Stellar Evolution \& Feedback Together \linebreak
\large
\linebreak
Summary of proposals from the Lorentz Center Workshop, 2022 \linebreak
\normalsize
 
\textbf{Co-authors:} (names and institutions)
  \linebreak
Sam Geen$^{1,2}$  \orcid{0000-0002-3150-2543},
Poojan Agrawal$^{3}$\orcid{0000-0002-1135-984X},
Paul A. Crowther$^{4}$\orcid{https://orcid.org/0000-0001-6000-6920},
B.W. Keller$^{5,18}$ \orcid{0000-0002-9642-7193},
Alex de Koter$^{1,6}$  \orcid{https://orcid.org/0000-0002-1198-3167}, 
Zsolt~Keszthelyi$^{1,7}$ \orcid{0000-0001-9663-9068}, 
Freeke van de Voort$^{8}$ \orcid{0000-0002-6301-638X},
Ahmad A. Ali$^{9}$ \orcid{0000-0001-5189-4022},
Frank Backs$^{1}$\orcid{0000-0003-3670-3181},
Lars Bonne$^{24}$\orcid{0000-0002-0915-4853},
Vittoria Brugaletta$^{10}$\orcid{0000-0003-1221-8771},
Annelotte Derkink $^1$\orcid{0000-0003-4627-5379},
Sylvia Ekstr\"om $^{11}$\orcid{0000-0002-2564-5660},
Yvonne A. Fichtner$^{12}$\orcid{0000-0003-4993-7200},
Luca Grassitelli$^{12}$,
Ylva G\"{o}tberg$^{23}$\orcid{0000-0002-6960-6911}, 
Erin R. Higgins$^{13}$ \orcid{0000-0003-2284-4469},
Eva Laplace$^{14}$\orcid{0000-0003-1009-5691},
Kong You Liow$^{9}$ \orcid{0000-0001-9309-1594},
Marta Lorenzo$^{15,27}$\orcid{0000-0002-4526-2018},
Anna F. McLeod$^{16,18}$ \orcid{0000-0002-5456-523X},
Georges Meynet $^{11}$\orcid{0000-0001-6181-1323}, Megan Newsome$^{25,26}$
G. Andr{\'e} Oliva$^{18}$\orcid{0000-0003-0124-1861},
Varsha Ramachandran$^{19}$\orcid{0000-0001-5205-7808},
Martin P. Rey,$^{20}$ \orcid{0000-0002-1515-995X},
Steven Rieder$^{11}$ \orcid{0000-0003-3688-5798}, 
Emilio Romano-D\'{\i}az$^{12}$\orcid{0000-0002-0071-3217}, 
Gautham Sabhahit$^{13}$\orcid{0000-0002-7442-1014},
Andreas A.C. Sander$^{19}$ \orcid{0000-0002-2090-9751},
Rafia Sarwar$^{21}$\orcid{0000-0002-6030-5741},
Hanno Stinshoff $^{10, 21}$\orcid{0000-0002-0514-1676},
Mitchel Stoop$^{1}$\orcid{0000-0003-4723-0447},
Dorottya Szécsi$^{21}$\orcid{0000-0001-6473-7085}, 
Maxime Trebitsch $^{22}$\orcid{0000-0002-6849-5375},
Jorick S. Vink$^{13}$\orcid{0000-0002-8445-4397},
Ethan Winch$^{13}$ 
\linebreak
(Author contact details and full list of institutions at end of paper)

\hfill


\hfill

\textbf{Abstract:} Stars strongly impact their environment, and shape structures on all scales throughout the universe, in a process known as ``feedback''. Due to the complexity of both stellar evolution and the physics of larger astrophysical structures, there remain many unanswered questions about how feedback operates, and what we can learn about stars by studying their imprint on the wider universe. In this white paper, we summarize discussions from the Lorentz Center meeting `Bringing Stellar Evolution and Feedback Together' in April 2022, and identify key areas where further dialogue can bring about radical changes in how we view the relationship between stars and the universe they live in.


\section{Introduction on Scales: From the Birth of Stars to the Wider Universe}
\begin{figure}[htbp]
    \centering
    \includegraphics[width=\textwidth]{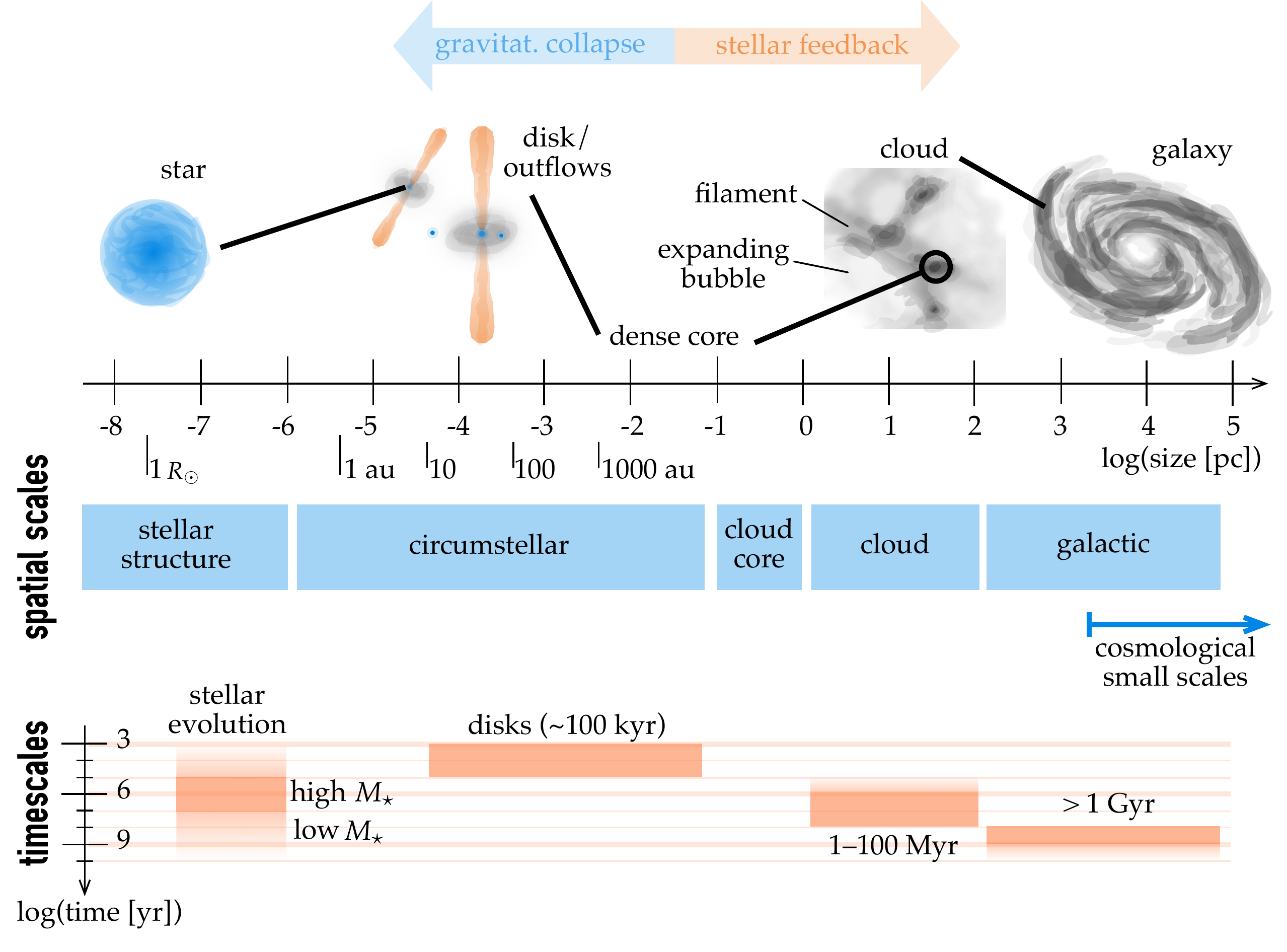}
    \caption{The different length scales of star formation in log-parsec. }
    \label{fig:small_large}
\end{figure}

Astrophysics spans many orders of magnitude in both physical distances and time. Researchers from different fields have varying definitions for what are considered ``small'' and ''large'' scales. Typically, ``small'' refers to processes smaller than those typically resolved in studies, whether observational or theoretical. Meanwhile, ``large'' typically refers to scales outside the boundaries of the problem domain. In Figure \ref{fig:small_large} we show a diagram depicting the range of relevant spatial and temporal scales, from stars to galaxies and beyond, in order to define and motivate discussions around the boundaries of domains of study considered in this work. 

The galactic scale, i.e. the largest physical scale considered here below the ``cosmological'' scale, is about 1 -- 100s of kpc.
A spiral galaxy like our Milky Way contains many (giant) molecular clouds of length scale 10 -- 100 pc, which from their dense cores can form star clusters at scales of 0.1 -- 10 pc.
Within those dense cores, the gravitational collapse that results in the formation of individual stars takes place.
Protostars are typically surrounded by accretion disks of sizes that range between 1 -- 1000 au, and outflows.
On the smallest physical scales considered here, we can regard the (intra)-stellar structure.
Within the star itself, we have the nuclear burning in the core, convection zones, envelope and stellar surface at 0.1 -- 10 $R_{\odot}$.

In numerical simulations, the connection between small and large scales is crucial because it is computationally expensive to set up and perform simulations that encompass the whole range of scales relevant to astrophysics within a reasonable amount of computing time. Despite this, an understanding of how the scales couple is important. various physical processes connect the smallest and largest scales with flows moving to both smaller and larger scales, often driven by the action of stars, in a cycle of material termed ``feedback''.

During the star formation process at stellar scales, the outflows launched by the disk and jet can influence the surrounding material. Ionizing radiation, stellar winds and eventual supernovae produced by the massive stars shape their natal molecular clouds and the interstellar medium, impacting subsequent generations of star formation. In this work we focus primarily on processes from stars after their formation phase ends, although protostellar outflows can be important both in themselves \citep{Federrath2014} and in concert with other feedback processes \citep{Kuiper2018} as stars form in molecular clouds \citep{Grudic2022,Verliat2022}. Feedback processes often act in concert, e.g. in the case of supernova feedback efficiency increasing if dense star-forming environments are dispersed by pre-supernova feedback \citep{Geen2015,lucas2020}.

Several techniques have been developed to bridge the different length scales.
From larger to smaller scales, zoomed-in simulations are performed, such that the regions from larger scale simulations are taken as initial conditions and the resolution of the regions is enhanced \citep[e.g.][]{Carlberg2022, Dobbs2022, Rey2022}.
This allows the regions of interest to be followed and studied more closely.

For example, zoom-in simulations of dense cloud cores can be used to follow their gravitational collapse into individual stars.
On the other hand, prescriptions are used to import the physics of smaller scales to the larger scales \citep[e.g.][]{Gutcke2021}.
This is generally done using empirical relations, analytical solutions, or parametric tables.
Some recent simulations employ multiple techniques to bridge the different scales \citep[e.g.][]{Rieder2022}. 


Critical tasks for the useful presentation and communication of the results of numerical simulations are: the determination of reliable intervals where a given quantity is valid or expected (e.g., the densities or angular momentum content of dense cores expected from simulations at the cloud scales), and the expression, whenever possible, of results that impact neighbouring scales using analytical formulae so that they can be used as prescriptions (e.g., evolutionary tracks for protostars that are used in larger-scale simulations).

With the advance of observational sites (e.g. Extremely Large Telescope, James Webb Space Telescope, Athena) with higher angular resolution, we come closer to resolving astrophysical structures large and small scales for regions in the Local Group and beyond.
Many of these sites will be able to resolve individual stars (of lower masses) for which before, we were only able to probe the large scale structures.
Observations and simulations of large and small scales in the (near) future will provide us with essential knowledge to connect these scales. 

    

\section{Introduction to Feedback: The Physics Connecting the Scales}

Once the protostellar phase has ended, stars impact their surroundings in a number of ways. We highlight some of the key processes by which stellar evolution processes drive feedback into the interstellar medium and beyond.

\subsection{Stellar Winds}\label{winds}

Stellar winds refer to the ejection of matter from a star's surface driven by radiation pressure on the gas in the star's atmosphere. Stellar winds impact their surroundings through a mixture of the mass loss rate $\dot{M}$ and terminal velocity $v_w$, i.e. the velocity that the stellar wind reaches once it is fully accelerated by radiation pressure. 

Observations by \cite{Groenewegen1989}, \cite{Prinja1990}, \cite{Crowther2016} and others confirm that these winds leave massive stars with terminal velocities that exceed 1000 km/s. This shocks the gas around the star to millions of degrees Kelvin, creating hot bubbles that drive strong flows into the interstellar medium \citep{Weaver1977}.

The rate of deposition of kinetic energy of stellar winds, $1/2 \dot{M} v_w^2$, is an important quantity in stellar feedback, where the energy in the wind bubble accumulates over time \citep{Weaver1977}. In the mode where stellar winds cool efficiently through thermal conduction or, more plausibly, turbulent mixing \citep[e.g.][]{Lancaster2021a}, the momentum deposition rate $\dot{M} v_w$ becomes more important. This mode is considerably weaker at driving large-scale flows since stored energy is lost. We examine in further detail how stellar wind bubbles impact nearby star-forming regions in Section \ref{local_group}. 

The properties of stars play a crucial role in setting $\dot{M}$ and $v_w$ \citep{Puls2015}. Factors such as metallicity \citep{Vink2001}, rotation \citep{Cranmer1995}, clumping \citep{Puls2008} and magnetic fields \citep{ud-Doula2002} are thought to play an important role in setting the precise wind properties. We return to these processes in detail later in the paper.

One of the most important stellar properties for determining $\dot{M}$ and $v_w$ is stellar mass. At solar metallicity, stars with masses larger than around 25$M_{\odot}$ do not make it to the cool red supergiant phase, but instead lose a lot of mass in line-driven winds \citep{Castor1975,Kudritzki2000,Vink2022}. At lower metallcity, winds become significantly weaker due to the lack of metal lines to couple radiation to the gas and drive material from the stellar surface.

A significant impediment to a better understanding of stellar winds is the uncertainty in mass loss rates. For stars below 25$M_{\odot}$, mass-loss rates are uncertain by 1-2 orders of magnitude in the so-called "weak-wind regime" \citep{Martins2005}. 

For those massive stars where mass-loss starts to dominate the evolution (at about 40$M_{\odot}$) the uncertainties are about a factor 2-3 \citep[e.g.][]{Bjorklund2021}. Such uncertainties were investigated in evolutionary models by \cite{Keszthelyi2017b}, finding that the discrepancies may be resolved by studying the rotational velocities of B-type supergiants \citep{Vink2010}, given that mass loss leads to angular momentum removal and spin-down of the stellar surface \citep{Langer1998,Maeder2000}.

Stars of order 80-100$M_{\odot}$ are in the transition region of \citet{Vink2012}, where mass-loss rates are known very accurately, but above this transition point, mass-loss rates included in most stellar evolution and population synthesis models are thought to be underestimated. 

\subsection{Ionizing Radiation}\label{LyC}
Stellar ionizing radiation can propagate and deposit energy on a large variety of scales, starting in the stars' own atmospheres and extending to the intergalactic medium across the Universe, where they ``reionized'' the universe after cosmic recombination. Pinpointing how much, and when, hard ionizing photons are released is thus a key input to model how stars affect their surroundings on all scales. We highlight here recent developments, open questions, and uncertainties in predicting the budget of ionizing photons from stellar evolution, and their coupling to galactic and intergalactic scales.
 
Ionizing fluxes of stars strongly depend on the star's temperature. Therefore, the fact that main-sequence stars are hotter at lower metallicities has a direct impact on the resulting ionizing photon budget. However, this effect could potentially be drastically or even totally altered by stellar evolution effects relating to rotation and binary interaction. Binary interaction can lead to mass exchange between the two stars, resulting in ``envelope-stripped'', and thus even hotter, helium stars. Rapid rotation is also thought to efficiently mix massive stars that cannot spin down at low metallicity, leading to the creation of helium-enriched, finally pure helium, stars, referred to as chemically homogeneous stars \citep{Yoon2005,Szecsi2015}. When determining the feedback for a resolved population of stars, it is therefore crucial to not miss the ``earliest'' (i.e.\ hottest) stars of the population as they dominate the ionizing feedback \citep[see, e.g.,][for recent examples]{Ramachandran2018b,Ramachandran2019}. In addition, accreting compact objects are known to emit X-rays and ionizing radiation which have been considered, to aid photoionization of interstellar or even intergalactic gas \citep{Chen2015,Schaerer2019,Senchyna2020}. 
Moreover,  cluster winds and superbubbles have recently been suggested as a source of additional ionizing flux \citep{oskinova2022}. While most of their emitted photons are too energetic to efficiently ionize gas, a fraction of them can contribute to the total budget of hydrogen and helium-ionizing photons in the universe. 

While the effective temperatures of stars can give some clues to their spectrum and ionizing power, black bodies only provide limited representations for the ionizing fluxes of hot stars. The absorption of radiation by recombination fronts inside the stellar wind can significantly reshape the spectral energy distribution, thereby considerably affecting the resulting quantities of ionizing photons emitted by the star. This is particularly striking for the He\,\textsc{ii} ionizing flux that is reduced by many orders of magnitude -- effectively vanishing -- if the stars manage to launch an optically thick (Wolf-Rayet type) wind \citep[e.g.][]{sander2020}. This effect is not an issue for hydrogen-ionizing photons, even though part of their flux budget is still consumed to drive stellar winds. 

Direct constraints of the ionizing flux of individual stars in the local Universe would be invaluable to constrain uncertainties of the sources of photoionization of interstellar gas, but is unfortunately limited by the unavailability of extreme UV (EUV) observational tools. Hence, other indirect methods are necessary, for example (1) inferring the ionizing emission from nebular spectra using scaling relations for recombination line luminosities, and (2) using the ionizing emission from computed stellar atmosphere models that sufficiently reproduce the spectrum at other wavelengths (UV, optical, IR). Since the stellar He\,\textsc{ii}-ionizing flux is considerably affected by winds from the star, UV observations remain an important tool to correctly determine the sources of these photons. 

Radiative feedback plays a key role in regulating the lifecycle of star-forming regions, and in providing an early mechanism to modify the phase and thermodynamics of gas in which massive stars then explode as supernovae to drive galactic outflows. The coupling between ionizing radiation, other sources of feedback and the surrounding gas however remains uncertain, due to the inherent challenges in modelling and observing these non-linear physical processes occurring on multiple spatial and time scales. Quantifying the balance between feedback budgets within H\,\textsc{ii} regions has now become possible \citep[e.g.,][]{lopez14, mcleod18, mcleod20, mcleod21, Olivier2021b, barnes20}. However, uncertainties pointed out above in stellar evolution and synthesizing stellar population outputs propagate into these measurements, making their interpretation challenging. Furthermore, the interaction between radiative, wind, and supernova feedback is a strongly non-linear process, which can lead to positive reinforcement and strong galactic outflow driving \citep[e.g.][]{lucas2020} or by contrast diminish the clustering of SN explosions and reduce their efficiency at expelling gas from a galaxy (e.g. \citealt{Agertz2020, Smith2021, Fichtner2022}). Pinpointing the sign and strength of these couplings, both observationally and theoretically will be key to interpreting galaxies in observations, understanding how they regulate their star-formation, how they enrich their surrounding environment in metals, and how radiation escapes from them to larger, cosmological scales.

H\,\textsc{i} reionization of the universe is mostly powered by stellar sources in low-mass star-forming galaxies \citep[e.g.][]{Robertson2015, Dayal2020, Yung2020, Trebitsch2021}, so having a good handle of their ionizing production is crucial, while keeping in mind that other sources of uncertainties (e.g. how much of this ionizing radiation escapes the ISM) still needs to be addressed.
Even prior to H\,\textsc{i} reionization, X-rays from the very early stellar populations in star-forming galaxies contribute to heating the IGM, but the rate of production of these X-rays is still uncertain. Most emission comes from X-ray binaries \citep[e.g.][]{Eide2018}, whose populations are poorly constrained at the highest redshifts. 21cm all-sky measurements are starting to put limits on the beginning of this heating era (\citealt{Bowman2018}), although other experiments are needed to confirm this result (see e.g. \citealt{Singh2022}). Next-generation facilities like the SKA will soon constrain the early heating of the Universe, making the need for detailed models timely. In this context, detailed understanding of binary evolution of stars (and in particular massive stars) is required to assess properly the early heating of the IGM.
While He\,II reionization, which happens at $z\sim 3$ \citep[e.g.][]{Worseck2016} is thought to be mostly dominated by AGN sources \citep[e.g.][]{Puchwein2019, FaucherGiguere2020}, the contribution from stellar populations remains mostly unconstrained. Notwithstanding the uncertainties on the escape fraction of He\,\textsc{ii}-ionizing photons, the uncertainties in the stellar population models pointed out above will translate to the contribution of these stellar populations to the He\,\textsc{ii} background. In particular, the presence of very massive stars or hydrogen-stripped stars (e.g. \citealt{Gotberg2020}) could strongly enhance the contribution of the overall stellar populations to He\,\textsc{ii} reionization.

\subsection{Supernovae}\label{SNe}
Feedback from supernovae (SN) has long been considered a key ingredient in studies of interstellar gas \citep[e.g.][]{McKee1977} and galaxy evolution \citep[e.g.][]{Larson1974}.  SNe, especially core-collapse Type II SNe, release significant $(\sim10^{51}\;\rm{erg})$ energy in the initial blastwave: sufficient to destroy molecular clouds \citep{White1991}, drive turbulence in the ISM \citep{McCray1979}, and power galactic winds and outflows \citep{Mathews1971}. These explosions are also major sources of metals, producing (for example) the vast bulk of interstellar oxygen \citep{Burbidge1957}.  Beyond core-collapse supernovae, thermonuclear (Type Ia) supernovae may also be a source of feedback energy, and also contribute to the cosmic metal budget \citep{Kawata2001}.  From the cloud- and galaxy-scale feedback perspective, the key questions connecting stellar evolution to supernovae feedback are as follows.  {\it Which stars will end their lives as supernovae?  When will these stars detonate their supernovae?  What will be the energy, mass, and metal returns of these supernovae events (and which form will the energy take at larger scales - kinetic or thermal)?}  Traditionally, very simple assumptions have been made about these questions:  all stars above a certain mass ($5-10\;\rm{M_\odot}$) detonate, with each ccSNe event depositing $\sim10^{51}\;\rm{erg}$ of energy and $\sim7-100\;\rm{M_\odot}$ of mass into the surrounding ISM \citep[e.g.][]{Katz1992}.  It has long been assumed that, at least on galactic scales, uncertainties in how this energy propagates through the ISM dominates over any uncertainties in stellar evolution models \citep{Naab2017,Rosdahl2017}, and that questions relating to the details of ccSNe detonation are swamped by uncertainties in the cooling and mixing rates of SN remnants.  However, recent studies \citep{Keller2022} and higher-resolution simulations \citep{Gutcke2021} have begun to reveal that the details of stellar evolution can detectably manifest themselves on galactic scales. 

Temporal evolution of the stellar structure, subject to internal and surface physical processes described in Section~\ref{InternalStellarProcesses} will lead to a stellar structure for which internal pressure gradients at some point will no longer be able to withstand the force of gravity. Understanding these processes will allow us to ultimately answer the three key questions identified above.  Hydrodynamical models of SNe detonation predict that the occurrence of underluminous \citep[e.g.][]{Lovegrove2013} and hyperluminous \citep[e.g.][]{Woosley2006} supernovae may occur for certain combinations of initial stellar mass, metallicity, and rotation.  Adding to this is the strong theoretical predictions for ``islands of explodability'', where SN progenitors will either produce very weak SN or in some cases directly collapse to form black holes (BHs) with no significant energy return whatsoever \citep{Smartt2009,Horiuchi2014,Sukhbold2020}.  Recent theoretical studies of binary star interactions have found that the significant changes induced to both the surface and core structure also will impact which stars detonate, and the energy of the subsequent SN \citep{Mueller2019,Laplace2021,Vartanyan2021}.  Despite these theoretical uncertainties, it is highly likely that theoretical models of galaxy evolution have in general {\it over-}estimated the SN energy budget, though this recently may be changing \citep{Emerick2019,Gutcke2021}.  Better observational constraints are needed to begin pinning down the true budget of energy for SN feedback.

Observationally, determining the SNe budget for stars across the IMF is extremely challenging, owing to the difficult problem of connecting SNe progenitors to individual SN events.  Red Supergiants (RSG) constitute the most common SN-progenitor stage, during which the star may experience a type IIP/L explosion \citep{Smartt2009}. However, the RSG phase may last $\sim 2.5 \times 10^{6}$ to $3 \times 10^{5}$\,yrs for stars ranging in initial mass between 9 and 20 $\rm{M_{\odot}}$ \citep{Meynet2015}, more massive stars, at high metallicity at least, potentially suffering from such intense mass loss that the entire envelope is lost and the stars first become yellow or blue supergiants before experiencing core collapse \citep[e.g.,][]{Graefener2016,Kee2021}. 
At lower metallicities, higher mass supergiants may exist and explode as e.g. pair-instability supernovae ejecting a peculiar chemical yield \citep{MartinezGonzalez2022}.
The RSG Betelgeuse experienced an unprecedented dimming of its visual brightness from December 2019 until April 2020, speculated to forewarn an imminent core-collapse. Though it appears that this event likely reflected a combination of surface activity and dust formation in a previously ejected gas cloud positioned in the line of sight \citep{Montarges2021}, the need for a dedicated monitoring campaign of a population of RSG stars for unexpected variability is clearly opportune and may help to identify systems for which an explosion may happen within about a human lifetime.  Alternatively, the collapse of such massive stars may lead to direct black hole formation with no or only little ejecta being expelled, consequently, with a very faint or undetectable supernova.  The most promising candidate for a disappearing star directly collapsing into a black hole showed evidence for an estimated $\sim0.5\;\rm{M_\odot}$ of ejecta \citep{Gerke2015, Sukhbold2020, Basinger2021}.  Wolf-Rayet stars, evolved stars that have lost or have been stripped from their hydrogen rich envelopes are alternative candidates for an impending Ib/c (or gamma-ray-burst) supernova explosion \citep[e.g.,][]{Groh2013a}. Within this group, Wolf-Rayet Oxygen (WO) stars are thought to be particularly evolved and in a post core-helium burning phase of evolution where timescales until core collapse are down to a few times $\sim 10^{3}$ or $10^{4}$\,yrs \citep{Meynet2015}. So far, only nine WO stars are known, the one thought to be closest to ending its life being WR102 with $\sim 1500$\,yrs left.  Other post-main sequence objects have been suggested as potential SN-progenitors, including Luminous Blue Variable (LBV) stars \citep{Kotak2006,Groh2013b} and Wolf-Rayet Nitrogen (WN) stars \citep{Groh2013a}. The former possibility is supported by evidence that the progenitor of SN 2005gl was possibly an LBV star \citep{GalYam2009}.

\subsection{Chemical Enrichment}\label{Chemical_Enrichment}

Nuclear processed material may be ejected from the star/system, and thus influence the chemical abundance of the surroundings, via at least three mechanisms; (i) stellar winds; (ii) supernova ejecta (discussed in detail in Sect.~\ref{SNe}); (iii)~(non-conservative) binary interaction (discussed further in Sect.~\ref{binaries}).
Consequently, whether nuclear processed material ends up in the interstellar medium after being created inside a star is a complex question. For example, elements that stay inside the star for a longer time (due to not being immediately ejected in the wind) may be able to undergo further nuclear processing. In the same way, elements may be ``saved'' by the wind from being processed further.
This makes the topic of chemical evolution a highly complex area of research with a number of impediments to our understanding of it.

\begin{figure}[htb]
\centering
\includegraphics[width=0.8\textwidth]{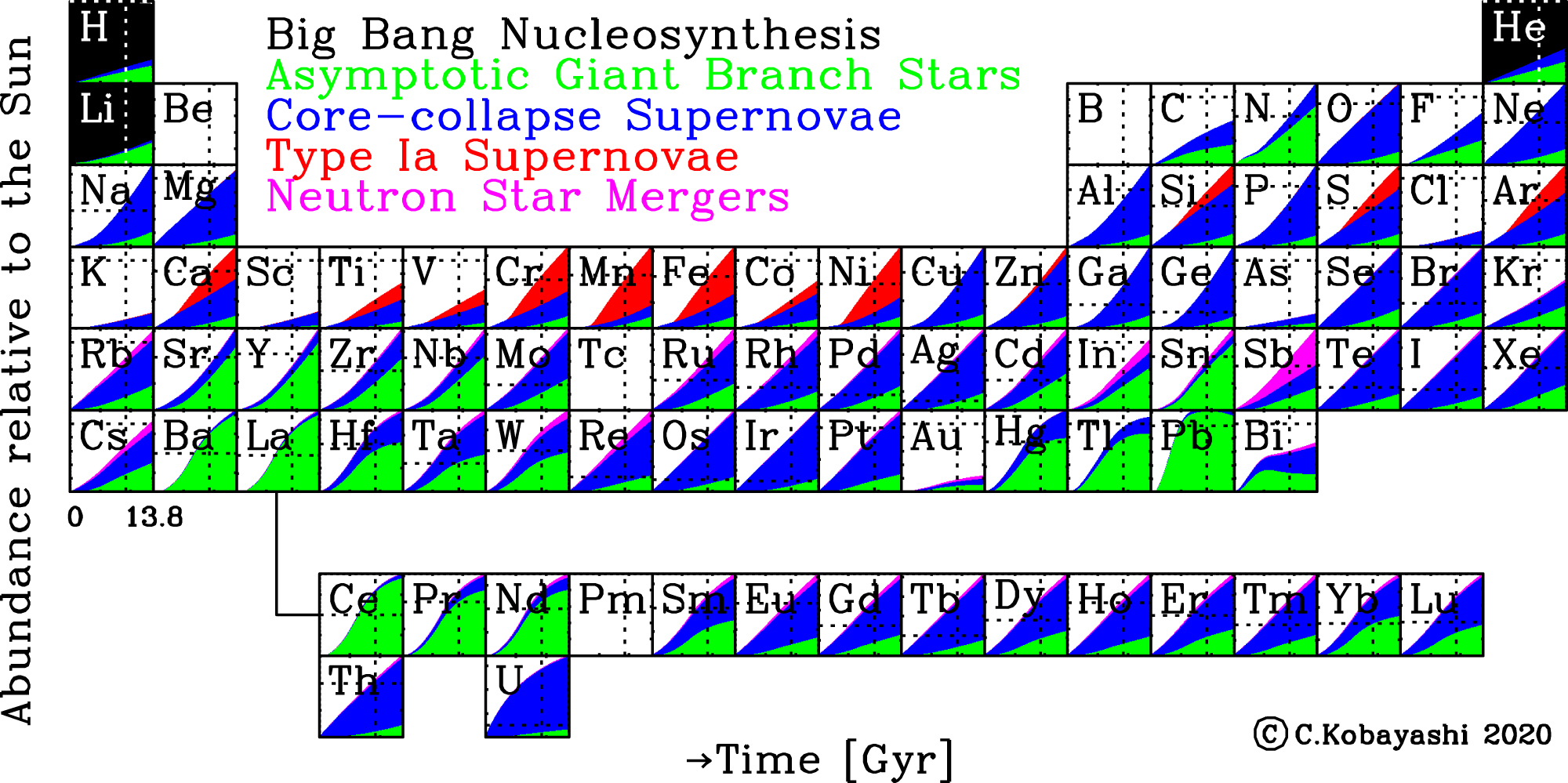}
\caption{Abundances relative to the Solar value plotted over time (in Gyr) for all the elements in the periodic table. This enables the reader to follow the different ways for evolution of the elements to take place via various processes. These processes include Big Bang Nucleosynthesis, AGB stars, Core-collapse Supernovae, Type Ia Supernovae and Neutron star mergers. Observations are depicted as dotted lines. From \cite{kobayashi2020}, reproduced with permission.}\label{fig:periodic}
\end{figure}

A deeper understanding of how and on which timescale elements are released in the interstellar medium is of great importance for modern stellar feedback simulations. Elements being ejected during the entire lifetime of a massive star could determine a different chemical evolution in the surrounding gas compared to the case in which they are “instantaneously” ejected in the supernova explosion. If we had a clearer view of these processes, we could also model more accurately how this enriched material spreads to larger scales, meaning the interstellar medium and the rest of the galaxy, because of turbulence and other mixing processes. 

Another important aspect in this regard is the comparison of the timescale in which the mixing of the newly-enriched material occurs in the gas with that of star formation. Will the mixing be fast enough to make the metallicity of the medium almost uniform, before a second generation of massive stars is born? As stars inherit their initial metallicity from the gas they have formed in, understanding how the timescales for chemical evolution and mixing relate to the time needed to form a new generation of stars would help to better understand their future evolution. Moreover, all these processes could be very different in low-metallicity environments, for which further analysis is recommended (see Section~\ref{local_group}).

The efficiency of mass-loss through stellar winds is highly dependent on the mass of the star (Sect.~\ref{winds}). The higher the mass, the higher the core temperature, leading to the activation of specific nuclear reactions. Massive and intermediate mass stars are known to have strong enough winds to eject nuclear-processed material. In particular, Asymptotic Giant Branch stars (AGBs) are important contributors to carbon and nitrogen via convective dredge up of nuclear products from the stellar core \citep{Romano10}. 


For the stellar wind (or interactions with a companion star) to be able to remove nuclear burning products, these products -- originally created in deep, hot burning regions -- need to already be found at the stellar surface. This can happen in two ways. Either the mixing between the deep layers and the surface needs to be strong (see Section~\ref{mixing}); or the layers from the top need to be first removed so the deeper layers are uncovered (see Section~\ref{winds}). In particular, mixing induced by rotation (or rotational mixing) has been shown to lead to extremely well mixed stars which evolve (quasi-)chemically homogeneously \citep{Maeder87}. But in less extreme cases, mixing (not only by rotation) can help bringing deeper layers upwards, to be lost in the wind eventually.
The decay of some isotopes serves as a counter to this process. This can be seen in the case of $^{26}$Al, which decays rather quickly (around 6\,s) into $^{26}$Mg (cf. \citealt{Finlay2012Al26}).

Figure~\ref{fig:periodic} shows the elements in the periodic table together with their cosmic origin \citep{kobayashi2020}. While the figure shows the state-of-the-art of our current knowledge, other possible avenues for the generation of elements are thought to exist. For example, gold has been proposed to form in kilonovae \citep{Kasen17}. 


Subsequent generations of stars have enriched interstellar gas with nuclear-processed elements. However, chemical enhancement is not only a time-dependent process but can be spatially traced as well. For example, the Milky Way displays a metallicity gradient \citep{Peimbert78,Afflerbach97} which decreases outwards, but other galaxies show other trends. 


Another source of uncertainty is the discrepancy between the yields found at the scale of stellar evolution modelling and those calculated at larger scales. To connect these two quantities, investigations are required with a varying degree of resolution as well as an understanding of the uncertainties involved in both calculations. Uncertainties include mixing and convection for single stars, tidal effects for binaries, and in general the handling of the Eddington limit. As one can see for example in \citet{Agrawal2022}, different approaches with multiple codes can lead to different predictions. Tracers such as CNO abundances may help resolve these discrepancies.

\section{Internal Stellar Processes}
\label{InternalStellarProcesses}

Stars are places where the four fundamental \rev{forces} in physics interact (viz., gravitational, electromagnetic, strong, and weak nuclear forces). Most global properties of stars can be inferred from the stellar structure equations, with the assumption of hydrostatic equilibrium. However, there are several key quantities e.g., nuclear reaction rates and opacity measurements (especially Iron or Fe), and internal processes in stars e.g., convection and overshooting, that remain highly uncertain in modelling stars, especially massive stars. 
Moreover, building accurate stellar models requires including the contribution of hydro- and magneto-hydrodynamical processes in the stellar interior such as stellar pulsations, stellar rotation and magnetic fields. These processes are not so well-understood and remain highly approximated in stellar models. 

Despite the recent progress in these areas in the last decade, several challenges remain in stellar physics. These include treatment of convection and the determination of the sizes of the convective zones, a proper account of all the processes that can induce mass loss at the different phases of evolution, the instabilities triggered in radiative zones that can transport angular momentum and chemical species (some of them likely triggered by rotation), and the impact of magnetic field in stellar interior and at the surface. Each of these uncertainties can severely impact stellar outputs and alter the feedback they inject into the interstellar medium. Below we discuss two significant internal processes.

\subsection{Internal Mixing}\label{mixing}

Energy produced in stars due to nuclear burning and other processes needs to be transported away to outer layers. The three main mechanisms responsible for this process are convection, conduction and radiation.  
In most stellar evolution codes, convection is modelled using a simple but successful formalism called mixing length theory \citep[MLT;][]{BohmVitense1958}. If energy is carried through convection, then owing to the actual movement of particles in the star, angular momentum and chemical species are also transported within the star. This can change the stellar structure and radius, which in turn affects the ionization, mass-loss rates and pre-supernova structure of the star \citep{Dessart:2013, Kaiser:2020}.

Convective boundary mixing (CBM) dictates the extension of the convective core and shell burning regions. There are multiple methods of implementing CBM with various mixing profiles such as core overshooting via step, exponential, or convective entrainment \citep{Scott2021}. The extension of the convective core via overshooting during core H-burning has various consequences leading to stars evolving at higher luminosities with increased mass loss over the integrated main sequence lifetime. Together, convection and associated mixing mechanisms contribute to the internal mixing in stars. 

Mixing processes can alter energy transport and the hydrogen content in the envelope, driving the evolution of massive stars towards red and blue supergiant phases and thus dictating red to blue supergiant ratios \citep{Schootemeijer2019}.
On the main sequence, the effects of internal mixing and mass loss dominate the evolutionary pathways which govern the fates of massive stars towards forming black holes and neutron stars. In the mass range $\sim$ 8--30 $M_{\odot}$\, interior mixing processes dominate the lives of massive stars, and in the mass range $\sim$ 30--60 $M_{\odot}$\, stellar winds drive the evolution towards Wolf-Rayet (WR) stripped Helium stars. The indirect effect of mass loss on interior mixing also plays a role in the switch of evolutionary path during core He-burning \citep{Higgins2020,Sabhahit2021}. The switch in evolutionary channels in post-MS evolution is key for predicting SNe progenitor populations.

Internal mixing mechanisms are one of the largest uncertainties in stellar physics. For example, the extent of core overshooting, which determines the length of the main sequence may itself be mass dependent \citep{Castro2014} which will also influence the post-main sequence evolutionary channels that form black holes. In fact, maintaining a sufficiently low core mass at the highest mass range can be critical in forming black holes and avoiding the pair instability supernovae regime \citep{Vink2021}. Similarly, radiative envelopes with subsurface convective layers can drive clumps in the wind, altering the mass-loss rates and having a large impact on SNe progenitors \citep{Davies2007,Cantiello2009,Jiang2015}, although there remain large uncertainties in these predictions.

Convection, as given by MLT, becomes highly inefficient in energy transport within the radiation dominated, low density envelopes of massive stars with $M_{\rm{init}}>40 M_{\odot}$ whose luminosities approaches the Eddington limit \citep[e.g.,][]{Langer1997,Maeder2009}, and only worsens for cooler supergiants owing to the hydrogen opacity bump at $T_\mathrm{eff} \sim 10^4 K$. Such a situation can cause stellar evolution codes to either crash or become stuck very small time-steps \citep{Paxton2013}. What happens in reality in such conditions, e.g., whether stars in close proximity to the Eddington limit inflate \citep{Grafener:2012} or not remains yet another unresolved problem. However stellar evolution models can predict widely different post-main sequence evolution when treating these highly inflated layers \citep{Agrawal2022}, which can have far reaching consequences in predicting the feedback properties of massive stars. Perhaps 2D or 3D simulations, or observational constraints such as the Humphreys-Davidson limit might shed light on what happens in such inflated, low density envelopes. 

Asteroseismology may provide calibrations for the efficiencies of internal mixing processes, but main sequence stars are usually fast rotators, and this can blur the period spacing. Low mass, slower rotators are more accessible for providing constraints with asteroseismology \citep{Pedersen2021,Bowman2021}. 
Rotation and rotational mixing play a major role in the enrichment of massive stars. The chemical enrichment of massive stars is dominated by rotational mixing instabilities, particularly whether the angular momentum is maintained via solid-body rotation, which is also important for determining neutron star spin.

\subsection{Stellar magnetic fields}
\label{B_fields}

Stars form in a magnetised medium, and recent simulations have demonstrated the large impact that magnetic fields play in the formation process (Oliva \& Kuiper, in prep.).
However, the acquisition of stellar magnetic fields is largely unconstrained.
There are two different kinds of magnetic fields that can be harboured by the massive stars. One possible branch is dynamos, either in the convective core driven by the $\alpha$-$\Omega$ cycle (similar to the surface of the Sun), or in the radiative layers driven by differential rotation \citep[e.g., the mechanism proposed by][]{Spruit2002}. Such dynamos are small-scale and vary on a short Alfv\'en timescale. In evolutionary models of massive stars, dynamo-generated magnetic fields in the radiative zones are commonly invoked \citep{Maeder2003,Maeder2004,Maeder2005,Heger2005,Potter2012,Fuller2019,takahashi2021}. 

Another branch of possibilities is relaxed, equilibrium fossil magnetic fields in the stellar radiative envelopes \citep[e.g.,][]{Braithwaite2004,Braithwaite2006}, which are large-scale and stable over the long-term evolution (Ohmic timescale). Such fields are now routinely observed via spectropolarimetry (exploiting the Zeeman effect) in a fraction of Galactic massive stars. Although no detections outside of the Galaxy have been made yet, largely due to the limitations of current instrumentation capabilities. 

The impact of fossil magnetic fields is far-reaching. These fields form a magnetosphere around the star, which channels the stellar outflow \citep{ud-Doula2002,Owocki2004}. 
The presence of magnetic fields can lead to two other important effects on mass loss: magnetic mass loss quenching (reducing the mass loss rate of the star, by up to an order of magnitude for a field of $\sim$ kG strength), and magnetic braking (removing angular momentum from the star and hence leading to an observable decrease of its surface rotation). Mass-loss quenching is a powerful mechanism that, independent of the metallicity, allows the star to retain most of its mass \citep{Georgy2017,Keszthelyi2017a,Keszthelyi2019,Keszthelyi2020,Keszthelyi2021,Petit2017}. 
The implementation of these processes in stellar evolution models has shown that magnetic braking very efficiently spins down the stellar surface and, depending on the internal coupling, may also produce observable surface nitrogen enrichment \citep{Meynet2011,Keszthelyi2019,Keszthelyi2021}, with a grid of stellar structure and evolution models available that take account of these processes \citep{Keszthelyi2022}.

Magnetic fields are thus a key component of stars. These are either built internally through internal dynamos or else retained as fossil fields from the time of the star's formation. While determining their presence and effect is difficult, recent advances can help us to better constrain and understand this problem.

\section{External Stellar Processes: Binaries}\label{binaries}

Similar to internal processes, external processes specific to the evolution of stars in multiple systems like tidal interactions, mass exchange, common envelope phases, stellar mergers can also impact the evolution and feedback of the stars. It is now established that binaries play a major role in the evolution of stellar populations \citep{Eldridge2020b,Eldridge:2022}. The majority of stars are born in binary or multiple systems and the binary fraction increases with stellar mass \citep{Moe:2017}. In addition, we now know that a significant fraction of these binaries will interact during their lifetime and initiate mass transfer, which has a significant impact on their structure and evolution \citep{Sana2012}. As a result of mass transfer, primaries can be stripped of their hydrogen envelope, which is accreted onto the secondary, spinning it up, or the system may merge. Consequently, their lifetimes and core properties change, affecting the final fate and stellar remnant. 

The picture is further complicated by the fact that both internal and external stellar processes, that are by themselves complex to properly model, can hardly be studied in isolation, as they all interact.
For example, stellar rotation, which can affect the evolution of stars, is strongly affected by tidal interactions in close binary systems. Indeed, tides can set up exchanges between two reservoirs of angular momentum, the orbital one and the rotational one, causing the star to spin-up or spin down depending on the circumstances and thus modifying the whole evolution of the two components by changing the rotation rates of the star and the radius of their orbits. A great diversity in evolutionary histories and stellar structures, for example at the time of core collapse, can be obtained through binary evolution. Likely some of the stellar pathways made possible by binary evolution are still to be discovered. 
Binary evolution impacts stellar feedback in three main ways: winds, ionizing radiation and supernovae rates.

\subsection{Impact on stellar winds}

The interstellar medium continuously receives mechanical energy and chemical feedback from stellar winds of the massive stars. Mass transfer in a close binary system will modify the nature of the wind from both components. The stripped primary (helium star) will likely possess a faster, lower density wind than its evolved (red supergiant) isolated counterpart, boosting the mechanical feedback. In addition, the mass-gaining secondary will usually produce a stronger wind as a result of its increased luminosity.

Helium stars (WR stars at high mass) contribute considerable energy to the total energy budget of a population \citep{Fichtner2022}. By way of example, in the SMC the collective wind of one multiple system (HD~5980) dominates over hundreds of OB stars in NGC~346. Stellar populations consisting of rotating stars in a binary system give raise to strong feedback processes specifically in low metallicities environment.

\subsection{Impact on the ionizing radiation}

It is well established that the ionizing radiation from a population of exclusively single (non-rotating) stars declines rapidly once the highest mass stars evolve off the main sequence, with a secondary (high energy) peak coinciding with the Wolf-Rayet phase \citep{Schmutz1992, Smith2002}. Since close binary evolution is capable of stripping the primary component of its hydrogen envelope, the effect of \rev{binary evolution} on the ionizing budget of young stellar populations is dramatic \citep{Gotberg2019}, especially at high energies (helium ionizing photons), and at low metallicities for which only exceptionally massive \rev{single} stars are capable of producing WR stars, \rev{whereas binary evolution leads to a prominent population of hot, stripped stars.}

 \citet{Rosdahl2018} found that, on average, binaries lead to escape fractions of $\sim$7–10 percent in the early universe, about three times higher than that produced by single stars only. With such a difference in ionizing escape fractions, their simulation of binary systems gives a cosmic reionization epoch before z$\sim$7, while the single-star escape fractions are not able to reionize their simulation volumes by z$\sim$6. Observationally, these findings have major implications for linking stellar evolution to cosmological-scale feedback. 


\subsection{Impact on core-collapse supernovae}

Binary evolution affects supernovae in three main ways: their energy budget, timing (location), and chemical yields. \citet{Zapartas2017} found that the inclusion of binaries in massive stellar systems substantially increases the number of supernovae expected among a stellar population, largely because of ``late" events originating from intermediate-mass ($4-8 M_{\odot}$) stars which would have otherwise evolved to white dwarfs, and whose binary interactions uniquely create the conditions for supernovae. The possibility of late events affects the delay-time distribution of supernovae: the maximum time expected for a single star to go supernova is 50 Myr, but late events occur on scales of $50-200$ Myr after birth. This stands in contrast with current prescriptions of supernovae timing in feedback simulations, which often assume an instantaneous explosion within 50 Myr for massive stars. 

Similarly, more massive stars that might otherwise be expected to collapse into black holes instead may experience mass stripping and common envelope interactions that create supernova conditions on the high-mass end as well. The widened range of initial masses that can experience supernovae from binary interactions will change the range of energetics expected and the properties of the supernova progenitors \citep[e.g.,][]{Podsiadlowski1992}. Moreover, mass transfer affects the structure and chemical composition of stars \citep[e.g.,][]{Laplace2021}, ultimately changing their chemical yields. For example, \citet{Farmer2021} showed recently that at solar metallicity, binary-stripped stars can eject twice as much carbon into their surroundings than single stars. In addition, binary systems can be the progenitors of gravitational wave sources, which are responsible for enriching stars in r-process elements \citep[][see also Sect.~\ref{Chemical_Enrichment}]{Kasen2017}. The supernova kick imparted at the moment of explosion of one binary component can result in a population of runaway and walkaways stars that explode in a location different from their birth environment \citep[e.g.,][]{Renzo2019}. 
    
\subsection{Impact of larger scales on binary formation}

Feedback processes in galaxies are thought to affect the formation of binaries and stellar multiples, through perturbations of gas clouds, feedback from stars and magnetic fields. Turbulence injected into molecular clouds through feedback from jets, winds and ionising radiation may affect when and how stellar multiples are formed. The quantity of angular momentum in protostar formation plays an important role in the mass of the protostellar disk, with more rotation leading to a more massive disk that fragments earlier. Bycontrast, if more mass is concentrated at the centre of the disk, a single massive star and/or a less massive companion will form. UV radiation and the propagation of heavy elements can also shape the formation of protostars as well as protoplanets.

Magnetic fields are important both in star-forming regions and also in stars (see Section \ref{B_fields}), and can play a role in coupling cloud scales to stellar scales. For example, sufficiently strong magnetic field will diminish fragmentation which then prevents but does not fully suppress binary formation. However, due to difficulties in resolution on a cloud-scale and the cost of small-scale simulations of protostar formation, simulations have not yet converged on the role that magnetic fields play in shaping in-situ binary formation.


 
Currently, most simulations do not generally take binary evolution into account in their feedback yields, however this is slowly changing in fields such as reionization studies \citep{Rosdahl2018} at $z > 6$, but recently in lower redshift galaxies such as \citet{Fichtner2022} for a sub-L* galaxy at $z=3$. 



\section{Varying Metallicity in our Local Group: The Effect of Z}
\label{local_group}


The Local Group is a complex environment with average present-day metallicities varying from $\sim 0.2~ Z_{\odot}$ in SagDIG \citep[][]{Saviane2002}, to $\sim2~Z_{\odot}$ in the Milky Way's Galactic Centre \citep[e.g.][]{NoguerasLara2018}. Additionally, significant metallicity gradients exist within galaxies \rev{\citep{Searle1971,VilaCostas1992,Henry1999}, including the Milky Way \citep[e.g.,][]{Lemasle2018}} - by metallicity of a galaxy, we typically refer to a radially averaged quantity. Stellar evolution and small-scale feedback models usually adopt the averaged values for a given galaxy when referencing their metallicities.

Within the Local Group, there are also large differences in densities and pressures, and star-forming mechanisms and rates. For example, the Large Magellanic Cloud hosts a million Solar-mass starburst region in 30 Doradus \citep[e.g.][]{Doran2013}, while Sextans A and the SMC appear to host isolated OB stars \citep[][]{Garcia2019, Lorenzo2022}. 

Our local universe thus presents a useful testbed for studying how stellar feedback operates in a variety of conditions. The role of metallicity applies to both the behaviour of stars themselves and the conditions in the gas in galaxies and hence shapes the interplay between the two (Brugaletta et al. in prep.).

In general, we assume that massive stars form with roughly the same metallicity as their local environment. Their surface abundances over their lifetime are shaped by chemical evolution as well as mixing and other processes such as envelope self-stripping, which drastically change the feedback properties of these stars. 

\subsection{Impact on Stellar Evolution and Feedback}

As discussed earlier, decreasing metallicity generally decreases the impact of stellar winds on an environment \citep{Vink2001}, since winds are driven by metal lines in the stellar atmosphere. This is largely a consequence of processes inside the star rather than the physics of the interstellar gas. Conversely, due to reduced photon absorption in the atmosphere, the ionizing photon emission rates are typically higher at lower stellar metallicity \citep{Martins2005}.



The effect on the gas around stars at lower metallicity is two-fold. The efficiency of mechanical and photoionization feedback is further enhanced by the fact that metal-line cooling in photoionized gas  \citep{Ferland2003} and collisionally-ionized gas \citep{Sutherland1993} is less efficient at low metallicity. However, lower dust fractions mean that the strength of radiation pressure decreases \citep{ali2021}.

The consequence of this on feedback depends on how these feedback processes couple, and if and when any given process dominates. Winds and supernovae create hot X-ray emitting bubbles ($10^6$ -- $10^8$ K), while photoionized regions are heated to $\sim10^4~$K. These regions co-exist within nebulae \citep{Guedel2008}, and their relative position and impact within feedback-driven nebulae remains a subject of active study. Analysis of observations in the Galactic Centre and compact H\,{\sc ii} regions shows that dust-processed radiation pressure dominates over other processes \citep{barnes20,Olivier21a}, while in the LMC/SMC/nearby galaxies, thermal pressure from photoionized gas dominates \citep{lopez14,mcleod18,mcleod21}. However, in addition to metallicity, these analyses are also affected by other environmental factors such as filling factors, ambient densities and pressures. Similarly, thermal losses are generally believed to have an important impact on wind bubbles in order to explain the missing energy in observed hot plasmas \citep{townsley03,lopez14}. These thermal losses may be more affected by turbulent mixing with cold gas in the environment of the wind bubble than by metal line cooling in the wind bubbles themselves \citep{Rosen2014,Lancaster2021a}.




\subsection{Low metallicity} 

There remain many unknowns concerning stellar evolution in extremely low metallicity environments due to the current limited observational capabilities and uncertain numerical ingredients, even in the case of single-star models. Depending on their metallicity, stars follow different evolutionary paths, resulting in different spectral subtypes dominating the mechanical and radiative yields. 
Between $\sim 1/10~$Z$_{\odot}$ and Z$_{\odot}$, the mechanical luminosity during stellar evolution is both theoretically and observationally expected to be dominated by Wolf-Rayet stars, despite their relatively short lifetimes and rarity \citep{Ramachandran2018, Fichtner2022}.
Instead, the more abundant stars with initial masses in the range $\sim$ 10-30~M$_{\odot}$ are expected to end their lives as SNe, hence dominate the mechanical luminosity after $\sim 10^7\,$yrs, i.e. at timescales comparable with the free-fall timescale of a young stellar cluster \citep{Krumholz2016}. 
At even lower metallicities, single-star evolution and wind models are not expected to lead to the appearance of the WR phenomenon, with the evolutionary channel leading to H-depleted stars being dominated by binary interaction \citep{Shenar20}. 

Their lower metal content may also lead to different evolutionary pathways that are not predicted at higher metallicities. Evolutionary models \citep{Brott2011} predict that, at metallicities lower than 1/10 $Z_{\odot}$, fast-rotating massive stars may evolve chemically homogeneously. In this evolutionary pathway, they can achieve temperatures hotter than the zero-age main sequence \citep{YoonLanger2005} and generally produce $\sim$ 5-10 times more ionizing energy than their normally-evolving counterparts \citep{Szecsi2015}.

The implications arising from the evidence that the majority of massive stars are in binary systems, and the lower angular momentum losses in low metallicity stellar models, are largely unconstrained. These effects are expected to attenuate the otherwise steeper decrease in kinetic energy feedback in the early phases of cluster formation at low metallicities \citep[][]{Fichtner2022}.  
However, the different evolutionary pathways do not only affect the yields estimated directly from evolutionary models. Stellar feedback, in fact, couples with the hydrodynamic evolution of the circumstellar gas. The slow and dense stellar outflows characteristic of cool supergiants are outside the line-driven regime and are only empirically constrained for stars in the Galactic Neighbourhood. It is likely that such slow gas can lead to thermal dissipation at sub-parsec scales, with a growing impact at low metallicities. Stars close to their Eddington limit during a Luminous Blue Variable phase (LBVs) are known to lose a significant fraction of their H-rich envelope during phases of high variability \citep{Humphreys1994,Vink2012}. Given the metallicity-independence of the HD limit \citep{Davies18, McDonald22}, and the higher expected number of redward-evolving stars at low-metallicities, one can expect that a larger fraction of the energy yield is dissipated well-before reaching the cluster scales \citep{Geen2015,Mackey15,Lancaster2021a}. Any systematic estimate must overcome our inability to convincingly model important stellar evolution phases such as the LBV phase \citep[however, see][]{Grassitelli21} and non-conservative mass-transfer phases in binary systems.

\section{Stars over Cosmic Time: The Effect of z}

In this Section we summarise discussions concerning how stellar evolution and feedback evolve over redshift. We focus our discussion here on redshifts up to $z\sim2$, the peak of cosmological star formation. There are likely to be significant differences between $z\sim2$ and very high redshift, in particular the role of the first (Population III) stars in the very early universe. As discussed earlier, aspects of stellar evolution such as binary evolution are likely to have a strong impact on cosmological processes such as reionization around $z \sim 6-11$.

Typical $z\sim2$ galaxies are moderately massive, deficient in iron-peak elements albeit $\alpha$/Fe enhanced \citep{Steidel2016}. Their nebular properties are relatively hard, and individual star forming knots (from lensing studies) indicate high star-formation intensities -- of order $\sim 0.1 M_{\odot}$/yr within a region of a few hundred parsecs \citep{Jones2010, Livermore2015}. Within the Local Group, only 30 Doradus (Tarantula Nebula) in the LMC displays such properties, albeit with a higher metallicity of $\sim 0.5 Z_{\odot}$ \citep{Crowther2019}.

\subsection{Star formation at low redshift ($z\sim0 - 0.3$)} 

Within the Local Group, where individual massive stars can generally be well spatially resolved, there are only a small number of actively star-forming galaxies whose current metallicity is $\leq 0.2 Z_{\odot}$, including the SMC, NGC 3109, IC 1613, Sextans A, WLM. Of these, the SMC has the highest star formation rate \citep{Kennicutt2008}, so is host to several hundred O stars, albeit with only a few dozen above 40 $M_{\odot}$ \citep{Schootemeijer2021}. Sextans A has an even lower metallicity \citep{vanZee2006} though also a lower star formation rate. In the context of star-forming knots at high redshift, these are modest, since such region will host thousands of O stars, hundreds of which are expected to exceed 40--50 $M_{\odot}$. The SMC and Sextans A therefore provide our only direct route to studying the evolution of massive stars at 0.1-0.2 Z{$_{\odot}$}, except at the highest masses, which are poorly sampled due to stochasticity. Sub-grid models employed in galaxy simulations (IMF, stellar models) are mainly constrained by local observations and then applied to simulations at high-$z$, or rely on theoretical predictions for low metallicity stars.

Metal poor massive stellar populations beyond the Local Group have been studied via integrated stellar populations, with the supergiant HII region Mrk 71 within NGC~2366 at 3 Mpc a striking example since it hosts massive super star clusters and has a metallicity of $\sim 0.15 Z_{\odot}$ \citep{GonzalezDelgado1994, Micheva2017}. This allows very massive metal poor stars to be observed at low metallicity, albeit in an integrated stellar population. In particular UV spectroscopy of the very young super star cluster Mrk 71-A with HST reveals strong HeII 1640 emission, providing a direct indicator of the presence of very massive stars (LJ Smith, priv. comm.). Mrk 71 is also notable in having evidence of leaking Lyman continuum photons \citep{Micheva2017}.

A sizeable population of Green Pea (GP) galaxies has been identified from SDSS observations whose properties overlap with high-redshift galaxies, i.e. both are metal-poor, possess high specific star formation rates plus hard nebular conditions in the BPT diagram \citep{Cardamone2009}, plus direct evidence for Lyman continuum leakage in some instances \citep{Izotov2016} and an excess soft X-ray emission \citep{Franeck2022}. In addition, there are examples of very metal-poor star forming galaxies locally with metallicities of only a few percent of the Solar Neighbourhood \citep[I~Zw\,18, SBS 0335][]{Lequeux1979,Izotov1990} which are potential analogues of star-forming galaxies in the very early Universe.
\citet{MadauDickinson2014} present the evolution of the average metal-content of the Universe through its history (their Fig. 14). For example, the metallicity of Sextans~A (1/10~Z$_{\odot}$) equates to $\sim$4~Gyr after the Big Bang.


\subsection{Star formation at $z\sim 2$}

Overall whilst there are some commonalities between metal-poor star forming regions locally and those at high redshift, some key differences remain, including composition \citep[Fe-poor, $\alpha$-enhanced,][]{Steidel2016}, higher specific star formation intensities potentially impacting on the IMF and close binary fraction, plus even if the mass and metallicity of a galaxy is the same at high- and low $z$, the environment, gas accretion and merger rate, AGN activity, will be different. 
It is speculated that old galactic globular clusters (GCs) in particular are born as Young Massive Clusters \citep[YMCs,][]{PortegiesZwart2010} from an $\alpha$-enhanced composition, with a first generation of metal-poor massive and intermediate-mass stars present \citep{Bastian2018} which could have contributed to the present-day chemical composition of the clusters \citep{deMink2009,Szecsi2018,Szecsi2019}. 

Regarding future prospects, efforts have recently been made to build extensive spectroscopic catalogues of massive stars in Local Group dwarf galaxies with sub-SMC metallicities \citep{Lorenzo2022}. These catalogues will yield a proper characterization of the physical parameters of metal-poor massive stars and will correct stellar evolutionary models. By introducing their physical properties as inputs of photoionization codes \citep[CLOUDY][]{Ferland1998}, we will be able to study the conditions of their surrounding interstellar medium and understand the stellar feedback of these metal-poor massive stars. Studying this interplay between individual massive stars and their surrounding interstellar medium in metal-poor environments can help us interpret the observations of high-$z$ galaxies and even estimate the amount of ionizing photons that dwarf galaxies contributed to the reionization of the Universe.

\section{From Star-by-Star Studies to IMF Averages and Population Synthesis}
The sources of feedback energy from massive stars -- their ionizing photon flux, the momentum carried by their stellar winds, and their ultimate fate as supernovae -- all depend strongly on the detailed physics of stellar evolution.   Without a clear understanding of the physical processes involved in the lives and deaths of massive stars, we cannot understand the ultimate impact of stellar feedback on galaxies.  Despite the urgency of this question, many theoretical studies of galaxy evolution make use of heavily simplified assumptions of how massive stars evolve.  How can we translate the best current understanding of stellar evolution into a better foundation for theoretical models of galaxy formation?

Stellar feedback in galaxies has been invoked as a mechanism to control the galactic star formation rate, the growth of spheroids, the baryon and metal content of galaxy discs, among other galaxy-scale properties.  Energy and momentum injected by massive stars can destroy star-forming clouds before they can convert the bulk of their gas into stars, and ultimately drive powerful galactic winds that remove baryons from the disc. Capturing these processes, either in semi-analytic models or hydrodynamic simulations, must begin with a robust budget (and timeline) of the relevant energy sources.

\subsection{What Matters at the scale of Galaxies?}
Broadly speaking, the primary physical process that makes galaxies ``care'' about the stellar populations they contain is feedback.  Galaxy-scale feedback is generally considered to be negative, with stellar feedback limiting galactic star formation by injecting turbulence \citep[e.g.][]{Padoan2016}, driving galactic outflows \citep[e.g.][]{Larson1974}, or destroying star-forming molecular clouds \citep[e.g.][]{Chevance2022}.  In addition to the energy and momentum that stellar populations inject into their surroundings, the mass-loss of stars can also pollute the interstellar medium (ISM) with metals produced in those stars, increasing the cooling rate of this gas and acting as a form of positive feedback \citep{Hirschmann2013}. Thus, the stellar physics that determines the energy and momentum of stellar winds, SN explosions, and UV radiation all act to change the impact of stellar feedback on the scale of galaxies.

For all but the smallest galaxies, the stellar populations driving feedback comprise tens of thousands or more stars.  In addition, simulations of galaxies typically cannot resolve individual stars except in the smallest, most isolated systems.  Thus, the primary questions that galactic astrophysicists must have for stellar astrophysicists come down to integrated or population-averaged quantities.  Simulations of galaxies may include supernovae, stellar winds, or UV feedback (or any combination of these). What is needed are mass loss, energy and momentum injection, and UV photon production rates as a function of time (in other words, yields of each of these quantities).  A detailed study of an individual star will not alone suffice for this: what is needed is an understanding of a fully-sampled IMF.  As the small-scale environment of individual stars is unknown and unresolved in these simulations, the only dependency of these quantities that can be probed are ones which are again population averaged, such as the birth metallicity \citep{Badenes2018} or ISM density\citep{Chabrier2014}.  The tool typically used to determine the population-averaged yields needed for galaxy simulations is Population Synthesis.

\subsection{Population Synthesis and Simple Stellar Populations}
No matter whether galaxies are modelled using analytic approximations, semi-analytic models, or full hydrodynamic simulations, the phenomena occurring inside and around individual stars necessarily must be averaged across large numbers ($10^3-10^7$) of stars.  Historically, this has been done through the use of Population Synthesis of Simple/Single Stellar Populations (SSPs).  SSPs are groups of stars, sampled from a given IMF \citep[e.g.][]{Leitherer1999}, that are assumed to have been born at a fixed time, with identical chemical properties.  Population synthesis models allow simulation codes to determine, as a function of time, the yields of mass, metals, and energy produced by the individual star particles within those simulations (or from an assumed population in an analytic or semi-analytic model).  Typically, this is done via either tabulated outputs from a population synthesis code \citep[e.g.][]{Leitherer1999,daSilva2012}, or through analytic functions fit to these yields.  While this hides much of the stellar physics involved in producing these yields ``under the hood'' of the population synthesis model, it does offer us the opportunity to easily incorporate more a sophisticated model of stellar evolution without significant work required to re-design galaxy simulation codes.

\section{Connecting Theory and Observations}

Theoretical approaches such as simulations are essential in astrophysics since laboratory experiments of most astronomical phenomena are impossible. Using theoretical results to inform observational results requires the creation of ``synthetic'' observations, or mock observational results generated using simulated inputs. This can take the form of simulated stellar spectra, multi-wavelength gas emission maps, mock galaxy catalogues, and more. This process is important both for observers, who may wish to understand the systems they observe with full 3D and time information, and theorists who wish to better constrain their models.

Creating mock observations is a complex process with many steps that must be treated properly to produce accurate results. This is a subject that has been widely discussed on various scales, from the regions around stars \cite[see review by][]{Haworth2018} to cosmological galaxy formation \cite[e.g.][]{Guidi2015}.

There are various hurdles relevant to stellar evolution and feedback that must be overcome if we are to close the gap between observed systems and theoretical predictions for how they behave. One key issue is ensuring that the physical structure of the observed system is realistic. This is highly affected by stellar feedback on all scales, which in turn is affected by the details of (massive) stellar evolution, as discussed in previous Sections. Conversely, with accurate theoretical models, it may be possible to use observations of feedback-driven structures as archaeological tools to inform studies of how stars evolve.

The motion of interstellar gas is chaotic, since it requires solutions to the coupled non-linear equations for (radiative magneto)hydrodynamics and N-body gravitation. This means that small perturbations to the early state of the cloud, such as initial seed turbulence or differences in stellar output, can have large cumulative effects on the later evolution of astrophysical systems. The variance from differences in stellar input and initial gas properties have been explored in star-forming regions \citep{Geen2018} and galaxies \citep{Keller2022}. Some linear response and mitigation of sampling errors is recoverable using statistical analysis and comparisons of large catalogues of both simulations and observations \citep{Eadie2018}. However, the physical divergence of solutions to sets of non-linear equations over time remains a serious concern in reproducing astronomical phenomena using simulations.

Simulations will often necessarily simplify or omit certain details of real-world physics for the sake of producing computationally-feasible or reducible results. Some models assume 1D or 2D geometries with symmetry in other dimensions, or ignore effects such as (non-)ideal magnetohydrodynamics, gas chemistry, thermal conduction, etc. Choices concerning simulated system size and resolution must also be made. Many of these assumptions may be reasonable and lead to minimal impact on the end result (e.g. through convergence in simulation resolution), but it is often hard to determine whether this is true without access to more expensive, physically-complete simulations.

Finally, the emission and absorption properties of stars and interstellar gas are complex, but are nonetheless required to be reproduced in detail if we wish to create accurate synthetic observations. This may be relatively simple for low-opacity systems with well-understood stellar populations, but becomes complex in other more general cases. Efforts have begun to connect the actions of stars to the emission properties of interstellar nebulae \citep[see, e.g.][]{Pellegrini2020}. However, the problem remains a difficult and costly one. A solution requires a good understanding of stellar evolution, feedback physics and gas microphysics and chemistry, all operating together over the lifetime of a system.

One mitigation to these problems may be found in posing questions in a way that reduces the impact of some of the uncertainties given above. Rather than producing a 1:1 comparison of individual objects, we may instead seek an interval of validity - that is to say, a set of possibilities informed by simulations that constrain certain parameters. Public data availability through standard databases would assist in this by allowing simulators and observers to access large quantities of relevant information, provided the limitations of the simulations and observations within the databases (e.g. resolution limits, systemic errors or important physical choices) are properly understood by the user. To ensure that the interval of validity and limitations are properly understood, increased collaborations between observers and simulators in the near future will be helpful.

\section{Conclusions}

\rev{The interplay between stars and their environment (termed ``stellar feedback'') is a long-standing problem that nonetheless is still the subject of active study. These questions remain open for numerous reasons, relating to the complexity of large-scale astrophysical gas dynamics and of the evolution of stars, individually and in multiple stellar systems.}

\rev{The outcome of the workshop was to identify a wide-ranging set of points of interaction between massive stars and the gas in galaxies, from the scale of protostellar disks to cosmological scales. In addition, the workshop highlighted the need for detailed discussions between researchers working on different aspects of both stellar evolution and feedback. For example, bridging the scales of molecular clouds and galaxies is important in tracking how the impact of massive stellar evolution is felt on (cosmological) galaxy scales.}

\rev{Much of this work is concerned with providing an inventory of the variables and unknowns affecting each field and how they relate to each other. For example, metallicity plays an important role in both the wind and radiation outputs from massive stars and the impact these processes have on the gas in galaxies through radiative cooling efficiencies. We provide detailed discussion of both theoretical and observed behaviour of stars and gas at different metallicities, using our local galactic environment and higher redshift galaxies as observational examples of this. Meanwhile, there remain strong uncertainties in the budget of mass, energy and chemical enrichment from winds, radiation and supernovae at different metallicities, including whether certain stars become supernovae at all (``islands of explodability'').}

\rev{We discuss the effects governing stellar evolution, including both internal effects such as mixing and magnetic fields, and external effects such as interaction with companion stars and how this shapes feedback. Determining the internal structure of stars remains difficult, although there are promising techniques for doing so using asteroseismology and comparison with theory, which in turn offers the ability to constrain a new generation of theoretical stellar evolution models. Multiple stellar evolution greatly complicates the evolutionary path of massive stars. Nonetheless, understanding stellar multiples remains crucial not only because a large fraction,  or even the majority, of massive stars are in binaries, but also because interacting binaries drastically change the feedback properties from massive stars, both before and after the stars go supernova. This in turn can even influence how cosmological processes such as reionization occur.}

\rev{We note that it is important to understand not just the action of individual stars or binary systems, but how feedback from stars combines as populations in galaxies. This in turn is important for determining what we know about individual stars when observing distant galaxies where individual stars cannot be resolved. }

\rev{Finally, we discuss efforts to compare theory and observations in detail. This remains a difficult task, since modelling the spectral emission from atmospheres of stars, as well as (photo and collisionally-)ionized gas is non-trivial, although more recently software tools are now able to perform this task. More worryingly, as (astrophysical) fluids evolve non-linearly and precise information about the initial state of an observed system is often difficult to obtain, direct one-to-one comparison is often challenging or impossible, and we must often rely on statistical comparisons.}

\rev{Overall, we believe that this is an exciting time to begin widening discussions between workers in the fields of stellar evolution and feedback, with advances in theory and observations in both fields allowing great improvements in our understanding of astrophysics, both from the point of view of the birth and evolution of stars in a galactic context, and also an inventory of how energy propagates from stars to shape local star formation, whole galaxies and the wider universe.}

\section{Acknowledgements}
\rev{We would like to thank the anonymous referee for their work in improving the quality of the manuscript.}
The workshop on which this manuscript is based was made possible thanks to the logistical and financial support of the Lorentz Center, Leiden, Netherlands. This funding is made available by Leiden University and the Dutch Science Foundation (NWO). The workshop was further supported by a NOVA grant for Star Formation, which SG also acknowledges as support. SG further acknowledges support from a Spinoza award of the NWO for research on the physics and chemistry of the interstellar medium. 
This research was partly funded by the National Science
Center (NCN), Poland under grant number OPUS 2021/41/B/ST9/00757. 
Y.A.F. and E.R.D. acknowledge support from Collaborative Research Center 956, sub-project C4, funded by the Deutsche Forschungsgemeinschaft (DFG) – project ID 184018867. Y.A.F was supported by the International Max Planck Research School in Astronomy and Astrophysics.
SR acknowledges funding from the European Research Council Horizon 2020 research and innovation programme (Grant No. 833925, project STAREX).
H.S. and D.Sz. were supported by the Alexander von Humboldt Foundation.
R.S was funded in part by the National Science Center(NCN), Poland under grant number OPUS 2021/41/B/ST9/00757. For the purpose of Open Access, the author has applied a CC-BY public copyright license to any Author Accepted Manuscript (AAM)version arising from this submission.
M.T. acknowledges support from the NWO grant 0.16.VIDI.189.162 (``ODIN'').
For the purpose of Open Access, the author has applied a CC-BY public copyright
license to any Author Accepted Manuscript (AAM) version arising from this
submission. 
A.A.C.S. and V.R. are supported by the Deutsche Forschungsgemeinschaft (DFG - German Research Foundation) in the form of an Emmy Noether Research Group -- Project-ID 445674056 (SA4064/1-1, PI Sander)"
M. L. gratefully acknowledges support by grants PID2019-105552RB-C41 and MDM-2017-0737 Unidad de Excelencia "María de Maeztu"-Centro de Astrobiología (CSIC-INTA), funded by MCIN/AEI/10.13039/501100011033 and  “ESF Investing in your future".

\hfill

\textbf{Contact:}

Name:	Sam Geen
 \linebreak						
Institution:  (1) Anton Pannekoek Institute for Astronomy, University of Amsterdam, 1098 XH Amsterdam, The Netherlands
(2) Leiden Observatory, Leiden University, PO Box 9513, 2300 RA Leiden, Netherlands
 \linebreak
Email: \href{mailto:s.t.geen@uva.nl}{s.t.geen@uva.nl}
 \linebreak

 \textbf{Full list of institutions:}
 
$^{1}$ Anton Pannekoek Institute for Astronomy, Universiteit van Amsterdam, Science Park 904, 1098 XH Amsterdam, Netherlands\\
$^{2}$ Leiden Observatory, Leiden University, PO Box 9513, 2300 RA Leiden, Netherlands \\
$^{3}$ McWilliams Center for Cosmology, Department of Physics, Carnegie Mellon University, Pittsburgh, PA 15213, USA\\
$^{4}$ Physics \& Astronomy, University of Sheffield, Hounsfield Road, Sheffield, S3 7RH, United Kingdom\\
$^{5}$ Department of Physics and Material Science, The University of Memphis, Memphis, TN 38152, USA\\
$^{6}$ Institute of Astronomy, KU Leuven, Celestijnenlaan 200D, 3001 Leuven, Belgium\\
$^{7}$ Center for Computational Astrophysics, Division of Science, National Astronomical Observatory of Japan, 2-21-1, Osawa, Mitaka, Tokyo 181-8588, Japan\\
$^{8}$ Cardiff Hub for for Astrophysics Research and Technology, School of Physics and Astronomy, Cardiff University, Queen's Buildings, The Parade, Cardiff CF24 3AA, UK\\
$^{9}$ Department of Physics and Astronomy, University of Exeter, Stocker Road, Exeter EX4 4QL, United Kingdom\\
$^{10}$ I. Physikalisches Institut, Universit{\"a}t zu K{\"o}ln, Z{\"u}lpicher Str. 77, 50937 Cologne, Germany\\
$^{11}$ Department of astronomy, University of Geneva, Chemin Pegasi 51, 1290 Versoix, Switzerland\\
$^{12}$Argelander-Institut f{\"u}r Astronomie, Universit{\"a}t Bonn, Auf dem H{\"u}gel 71, D-53121 Bonn, Germany\\
$^{13}$ Armagh Observatory \& Planetarium, College Hill, Armagh, BT619DG, United Kingdom\\
$^{14}$ Heidelberger Institut f{\"ur} Theoretische Studien, Schloss-Wolfsbrunnenweg 35, 69118 Heidelberg, Germany\\
$^{15}$ Centro de Astrobiología, CSIC-INTA. Crtra. de Torrejón a Ajalvir km 4. 28850 Torrejón de Ardoz (Madrid), Spain\\
$^{16}$Centre for Extragalactic Astronomy, Department of Physics, Durham University, South Road,  Durham DH1 3LE, United Kingdom\\
$^{17}$Institute for Computational Cosmology, Department of Physics, University of Durham, South Road, Durham DH1 3LE, United Kingdom\\
$^{18}$ Institute for Astronomy and Astrophysics, University of T{\"u}bingen, Auf der Morgenstelle 10, 72076 T{\"u}bingen, Germany\\
$^{19}$ Zentrum f{\"u}r Astronomie der Universit{\"a}t Heidelberg, Astronomisches Rechen-Institut, M{\"o}nchhofstr. 12-14, 69120 Heidelberg, Germany\\
$^{20}$ Sub-department of Astrophysics, University of Oxford, DWB, Keble Road, Oxford OX1 3RH, United Kingdom \\
$^{21}$ Institute of Astronomy, Faculty of Physics, Astronomy and Informatics, Nicolaus Copernicus University, Grudzi\k{a}dzka 5, 87-100 Toruń, Poland\\
$^{22}$ Kapteyn Astronomical Institute, University of Groningen, P.O. Box 800, 9700 AV Groningen, Netherlands \\
$^{23}$ The Observatories of the Carnegie Institution for Science, 813 Santa Barbara Street, CA-91101 Pasadena, USA \\
$^{24}$ SOFIA Science Center, USRA, NASA Ames Research Center, Moffett Field, CA 94045, USA\\
$^{25}$ Las Cumbres Observatory, 6740 Cortona Dr, Suite 102, Goleta, CA 93117-5575, USA\\
$^{26}$ Department of Physics, University of California, Santa Barbara, CA 93106-9530, USA \\
$^{27}$ Departamento de Física Teórica, Universidad Autónoma de Madrid (UAM), Campus de Cantoblanco, E-28049 Madrid, Spain

\bibliographystyle{mnras}
\bibliography{bseft}

\begin{thebibliography}{}
\makeatletter
\relax
\def\mn@urlcharsother{\let\do\@makeother \do\$\do\&\do\#\do\^\do\_\do\%\do\~}
\def\mn@doi{\begingroup\mn@urlcharsother \@ifnextchar [ {\mn@doi@}
  {\mn@doi@[]}}
\def\mn@doi@[#1]#2{\def\@tempa{#1}\ifx\@tempa\@empty \href
  {http://dx.doi.org/#2} {doi:#2}\else \href {http://dx.doi.org/#2} {#1}\fi
  \endgroup}
\def\mn@eprint#1#2{\mn@eprint@#1:#2::\@nil}
\def\mn@eprint@arXiv#1{\href {http://arxiv.org/abs/#1} {{\tt arXiv:#1}}}
\def\mn@eprint@dblp#1{\href {http://dblp.uni-trier.de/rec/bibtex/#1.xml}
  {dblp:#1}}
\def\mn@eprint@#1:#2:#3:#4\@nil{\def\@tempa {#1}\def\@tempb {#2}\def\@tempc
  {#3}\ifx \@tempc \@empty \let \@tempc \@tempb \let \@tempb \@tempa \fi \ifx
  \@tempb \@empty \def\@tempb {arXiv}\fi \@ifundefined
  {mn@eprint@\@tempb}{\@tempb:\@tempc}{\expandafter \expandafter \csname
  mn@eprint@\@tempb\endcsname \expandafter{\@tempc}}}

\bibitem[\protect\citeauthoryear{{Afflerbach}, {Churchwell}  \&
  {Werner}}{{Afflerbach} et~al.}{1997}]{Afflerbach97}
{Afflerbach} A.,  {Churchwell} E.,   {Werner} M.~W.,  1997, \mn@doi [\apj]
  {10.1086/303771}, \href
  {https://ui.adsabs.harvard.edu/abs/1997ApJ...478..190A} {478, 190}

\bibitem[\protect\citeauthoryear{{Agertz} et~al.,}{{Agertz}
  et~al.}{2020}]{Agertz2020}
{Agertz} O.,  et~al., 2020, \mn@doi [\mnras] {10.1093/mnras/stz3053}, \href
  {https://ui.adsabs.harvard.edu/abs/2020MNRAS.491.1656A} {491, 1656}

\bibitem[\protect\citeauthoryear{{Agrawal}, {Sz{\'e}csi}, {Stevenson},
  {Eldridge}  \& {Hurley}}{{Agrawal} et~al.}{2022}]{Agrawal2022}
{Agrawal} P.,  {Sz{\'e}csi} D.,  {Stevenson} S.,  {Eldridge} J.~J.,   {Hurley}
  J.,  2022, \mn@doi [\mnras] {10.1093/mnras/stac930}, \href
  {https://ui.adsabs.harvard.edu/abs/2022MNRAS.512.5717A} {512, 5717}

\bibitem[\protect\citeauthoryear{{Ali}}{{Ali}}{2021}]{ali2021}
{Ali} A.~A.,  2021, \mn@doi [\mnras] {10.1093/mnras/staa3992}, \href
  {https://ui.adsabs.harvard.edu/abs/2021MNRAS.501.4136A} {501, 4136}

\bibitem[\protect\citeauthoryear{{Badenes} et~al.,}{{Badenes}
  et~al.}{2018}]{Badenes2018}
{Badenes} C.,  et~al., 2018, \mn@doi [\apj] {10.3847/1538-4357/aaa765}, \href
  {https://ui.adsabs.harvard.edu/abs/2018ApJ...854..147B} {854, 147}

\bibitem[\protect\citeauthoryear{{Barnes}, {Longmore}, {Dale}, {Krumholz},
  {Kruijssen}  \& {Bigiel}}{{Barnes} et~al.}{2020}]{barnes20}
{Barnes} A.~T.,  {Longmore} S.~N.,  {Dale} J.~E.,  {Krumholz} M.~R.,
  {Kruijssen} J.~M.~D.,   {Bigiel} F.,  2020, \mn@doi [\mnras]
  {10.1093/mnras/staa2719}, \href
  {https://ui.adsabs.harvard.edu/abs/2020MNRAS.498.4906B} {498, 4906}

\bibitem[\protect\citeauthoryear{{Basinger}, {Kochanek}, {Adams}, {Dai}  \&
  {Stanek}}{{Basinger} et~al.}{2021}]{Basinger2021}
{Basinger} C.~M.,  {Kochanek} C.~S.,  {Adams} S.~M.,  {Dai} X.,   {Stanek}
  K.~Z.,  2021, \mn@doi [\mnras] {10.1093/mnras/stab2620}, \href
  {https://ui.adsabs.harvard.edu/abs/2021MNRAS.508.1156B} {508, 1156}

\bibitem[\protect\citeauthoryear{{Bastian} \& {Lardo}}{{Bastian} \&
  {Lardo}}{2018}]{Bastian2018}
{Bastian} N.,  {Lardo} C.,  2018, \mn@doi [\araa]
  {10.1146/annurev-astro-081817-051839}, \href
  {https://ui.adsabs.harvard.edu/abs/2018ARA&A..56...83B} {56, 83}

\bibitem[\protect\citeauthoryear{{Bj{\"o}rklund}, {Sundqvist}, {Puls}  \&
  {Najarro}}{{Bj{\"o}rklund} et~al.}{2021}]{Bjorklund2021}
{Bj{\"o}rklund} R.,  {Sundqvist} J.~O.,  {Puls} J.,   {Najarro} F.,  2021,
  \aap, 648, A36

\bibitem[\protect\citeauthoryear{{B{\"o}hm-Vitense}}{{B{\"o}hm-Vitense}}{1958}]{BohmVitense1958}
{B{\"o}hm-Vitense} E.,  1958, \zap, \href
  {https://ui.adsabs.harvard.edu/abs/1958ZA.....46..108B} {46, 108}

\bibitem[\protect\citeauthoryear{{Bowman}}{{Bowman}}{2021}]{Bowman2021}
{Bowman} D.~M.,  2021, in OBA Stars: Variability and Magnetic Fields. p.~27,
  \mn@doi{10.5281/zenodo.5109690}

\bibitem[\protect\citeauthoryear{{Bowman}, {Rogers}, {Monsalve}, {Mozdzen}  \&
  {Mahesh}}{{Bowman} et~al.}{2018}]{Bowman2018}
{Bowman} J.~D.,  {Rogers} A. E.~E.,  {Monsalve} R.~A.,  {Mozdzen} T.~J.,
  {Mahesh} N.,  2018, \mn@doi [\nat] {10.1038/nature25792}, \href
  {https://ui.adsabs.harvard.edu/abs/2018Natur.555...67B} {555, 67}

\bibitem[\protect\citeauthoryear{{Braithwaite} \& {Nordlund}}{{Braithwaite} \&
  {Nordlund}}{2006}]{Braithwaite2006}
{Braithwaite} J.,  {Nordlund} {\r{A}}.,  2006, \mn@doi [\aap]
  {10.1051/0004-6361:20041980}, \href
  {https://ui.adsabs.harvard.edu/abs/2006A&A...450.1077B} {450, 1077}

\bibitem[\protect\citeauthoryear{{Braithwaite} \& {Spruit}}{{Braithwaite} \&
  {Spruit}}{2004}]{Braithwaite2004}
{Braithwaite} J.,  {Spruit} H.~C.,  2004, \mn@doi [\nat] {10.1038/nature02934},
  \href {https://ui.adsabs.harvard.edu/abs/2004Natur.431..819B} {431, 819}

\bibitem[\protect\citeauthoryear{{Brott} et~al.,}{{Brott}
  et~al.}{2011}]{Brott2011}
{Brott} I.,  et~al., 2011, \mn@doi [\aap] {10.1051/0004-6361/201016113}, \href
  {https://ui.adsabs.harvard.edu/abs/2011A&A...530A.115B} {530, A115}

\bibitem[\protect\citeauthoryear{{Burbidge}, {Burbidge}, {Fowler}  \&
  {Hoyle}}{{Burbidge} et~al.}{1957}]{Burbidge1957}
{Burbidge} E.~M.,  {Burbidge} G.~R.,  {Fowler} W.~A.,   {Hoyle} F.,  1957,
  \mn@doi [Reviews of Modern Physics] {10.1103/RevModPhys.29.547}, \href
  {https://ui.adsabs.harvard.edu/abs/1957RvMP...29..547B} {29, 547}

\bibitem[\protect\citeauthoryear{{Cantiello} et~al.,}{{Cantiello}
  et~al.}{2009}]{Cantiello2009}
{Cantiello} M.,  et~al., 2009, \mn@doi [\aap] {10.1051/0004-6361/200911643},
  \href {https://ui.adsabs.harvard.edu/abs/2009A&A...499..279C} {499, 279}

\bibitem[\protect\citeauthoryear{{Cardamone} et~al.,}{{Cardamone}
  et~al.}{2009}]{Cardamone2009}
{Cardamone} C.,  et~al., 2009, \mn@doi [\mnras]
  {10.1111/j.1365-2966.2009.15383.x}, \href
  {https://ui.adsabs.harvard.edu/abs/2009MNRAS.399.1191C} {399, 1191}

\bibitem[\protect\citeauthoryear{{Carlberg} \& {Keating}}{{Carlberg} \&
  {Keating}}{2022}]{Carlberg2022}
{Carlberg} R.~G.,  {Keating} L.~C.,  2022, \mn@doi [\apj]
  {10.3847/1538-4357/ac347e}, \href
  {https://ui.adsabs.harvard.edu/abs/2022ApJ...924...77C} {924, 77}

\bibitem[\protect\citeauthoryear{{Castor}, {Abbott}  \& {Klein}}{{Castor}
  et~al.}{1975}]{Castor1975}
{Castor} J.~I.,  {Abbott} D.~C.,   {Klein} R.~I.,  1975, \apj, 195, 157

\bibitem[\protect\citeauthoryear{{Castro}, {Fossati}, {Langer},
  {Sim{\'o}n-D{\'\i}az}, {Schneider}  \& {Izzard}}{{Castro}
  et~al.}{2014}]{Castro2014}
{Castro} N.,  {Fossati} L.,  {Langer} N.,  {Sim{\'o}n-D{\'\i}az} S.,
  {Schneider} F.~R.~N.,   {Izzard} R.~G.,  2014, \mn@doi [\aap]
  {10.1051/0004-6361/201425028}, \href
  {https://ui.adsabs.harvard.edu/abs/2014A&A...570L..13C} {570, L13}

\bibitem[\protect\citeauthoryear{{Chabrier}, {Hennebelle}  \&
  {Charlot}}{{Chabrier} et~al.}{2014}]{Chabrier2014}
{Chabrier} G.,  {Hennebelle} P.,   {Charlot} S.,  2014, \mn@doi [\apj]
  {10.1088/0004-637X/796/2/75}, \href
  {https://ui.adsabs.harvard.edu/abs/2014ApJ...796...75C} {796, 75}

\bibitem[\protect\citeauthoryear{{Chen}, {Woods}, {Yungelson}, {Gilfanov}  \&
  {Han}}{{Chen} et~al.}{2015}]{Chen2015}
{Chen} H.-L.,  {Woods} T.~E.,  {Yungelson} L.~R.,  {Gilfanov} M.,   {Han} Z.,
  2015, \mn@doi [\mnras] {10.1093/mnras/stv1865}, \href
  {https://ui.adsabs.harvard.edu/abs/2015MNRAS.453.3024C} {453, 3024}

\bibitem[\protect\citeauthoryear{{Chevance} et~al.,}{{Chevance}
  et~al.}{2022}]{Chevance2022}
{Chevance} M.,  et~al., 2022, \mn@doi [\mnras] {10.1093/mnras/stab2938}, \href
  {https://ui.adsabs.harvard.edu/abs/2022MNRAS.509..272C} {509, 272}

\bibitem[\protect\citeauthoryear{{Cranmer} \& {Owocki}}{{Cranmer} \&
  {Owocki}}{1995}]{Cranmer1995}
{Cranmer} S.~R.,  {Owocki} S.~P.,  1995, \mn@doi [\apj] {10.1086/175272}, \href
  {https://ui.adsabs.harvard.edu/abs/1995ApJ...440..308C} {440, 308}

\bibitem[\protect\citeauthoryear{{Crowther}}{{Crowther}}{2019}]{Crowther2019}
{Crowther} P.~A.,  2019, \mn@doi [Galaxies] {10.3390/galaxies7040088}, \href
  {https://ui.adsabs.harvard.edu/abs/2019Galax...7...88C} {7, 88}

\bibitem[\protect\citeauthoryear{Crowther et~al.,}{Crowther
  et~al.}{2016}]{Crowther2016}
Crowther P.~A.,  et~al., 2016, \mn@doi [Monthly Notices of the Royal
  Astronomical Society] {10.1093/mnras/stw273}, 458, 624

\bibitem[\protect\citeauthoryear{{Davies}, {Vink}  \& {Oudmaijer}}{{Davies}
  et~al.}{2007}]{Davies2007}
{Davies} B.,  {Vink} J.~S.,   {Oudmaijer} R.~D.,  2007, \mn@doi [\aap]
  {10.1051/0004-6361:20077193}, \href
  {https://ui.adsabs.harvard.edu/abs/2007A&A...469.1045D} {469, 1045}

\bibitem[\protect\citeauthoryear{{Davies}, {Crowther}  \& {Beasor}}{{Davies}
  et~al.}{2018}]{Davies18}
{Davies} B.,  {Crowther} P.~A.,   {Beasor} E.~R.,  2018, \mn@doi [\mnras]
  {10.1093/mnras/sty1302}, \href
  {https://ui.adsabs.harvard.edu/abs/2018MNRAS.478.3138D} {478, 3138}

\bibitem[\protect\citeauthoryear{{Dayal} et~al.,}{{Dayal}
  et~al.}{2020}]{Dayal2020}
{Dayal} P.,  et~al., 2020, \mn@doi [\mnras] {10.1093/mnras/staa1138}, \href
  {https://ui.adsabs.harvard.edu/abs/2020MNRAS.495.3065D} {495, 3065}

\bibitem[\protect\citeauthoryear{{Dessart}, {Hillier}, {Waldman}  \&
  {Livne}}{{Dessart} et~al.}{2013}]{Dessart:2013}
{Dessart} L.,  {Hillier} D.~J.,  {Waldman} R.,   {Livne} E.,  2013, \mn@doi
  [\mnras] {10.1093/mnras/stt861}, \href
  {https://ui.adsabs.harvard.edu/abs/2013MNRAS.433.1745D} {433, 1745}

\bibitem[\protect\citeauthoryear{{Dobbs}, {Bending}, {Pettitt}  \&
  {Bate}}{{Dobbs} et~al.}{2022}]{Dobbs2022}
{Dobbs} C.~L.,  {Bending} T.~J.~R.,  {Pettitt} A.~R.,   {Bate} M.~R.,  2022,
  \mn@doi [\mnras] {10.1093/mnras/stab3036}, \href
  {https://ui.adsabs.harvard.edu/abs/2022MNRAS.509..954D} {509, 954}

\bibitem[\protect\citeauthoryear{{Doran} et~al.,}{{Doran}
  et~al.}{2013}]{Doran2013}
{Doran} E.~I.,  et~al., 2013, \mn@doi [\aap] {10.1051/0004-6361/201321824},
  \href {https://ui.adsabs.harvard.edu/abs/2013A&A...558A.134D} {558, A134}

\bibitem[\protect\citeauthoryear{{Eadie}, {Keller}  \& {Harris}}{{Eadie}
  et~al.}{2018}]{Eadie2018}
{Eadie} G.,  {Keller} B.,   {Harris} W.~E.,  2018, \mn@doi [\apj]
  {10.3847/1538-4357/aadb95}, \href
  {https://ui.adsabs.harvard.edu/abs/2018ApJ...865...72E} {865, 72}

\bibitem[\protect\citeauthoryear{{Eide}, {Graziani}, {Ciardi}, {Feng},
  {Kakiichi}  \& {Di Matteo}}{{Eide} et~al.}{2018}]{Eide2018}
{Eide} M.~B.,  {Graziani} L.,  {Ciardi} B.,  {Feng} Y.,  {Kakiichi} K.,   {Di
  Matteo} T.,  2018, \mn@doi [\mnras] {10.1093/mnras/sty272}, \href
  {https://ui.adsabs.harvard.edu/abs/2018MNRAS.476.1174E} {476, 1174}

\bibitem[\protect\citeauthoryear{{Eldridge} \& {Stanway}}{{Eldridge} \&
  {Stanway}}{2020}]{Eldridge2020b}
{Eldridge} J.~J.,  {Stanway} E.~R.,  2020, arXiv e-prints, \href
  {https://ui.adsabs.harvard.edu/abs/2020arXiv200511883E} {p. arXiv:2005.11883}

\bibitem[\protect\citeauthoryear{{Eldridge} \& {Stanway}}{{Eldridge} \&
  {Stanway}}{2022}]{Eldridge:2022}
{Eldridge} J.~J.,  {Stanway} E.~R.,  2022, arXiv e-prints, \href
  {https://ui.adsabs.harvard.edu/abs/2022arXiv220201413E} {p. arXiv:2202.01413}

\bibitem[\protect\citeauthoryear{{Emerick}, {Bryan}  \& {Mac Low}}{{Emerick}
  et~al.}{2019}]{Emerick2019}
{Emerick} A.,  {Bryan} G.~L.,   {Mac Low} M.-M.,  2019, \mn@doi [\mnras]
  {10.1093/mnras/sty2689}, \href
  {https://ui.adsabs.harvard.edu/abs/2019MNRAS.482.1304E} {482, 1304}

\bibitem[\protect\citeauthoryear{{Farmer}, {Laplace}, {de Mink}  \&
  {Justham}}{{Farmer} et~al.}{2021}]{Farmer2021}
{Farmer} R.,  {Laplace} E.,  {de Mink} S.~E.,   {Justham} S.,  2021, \mn@doi
  [\apj] {10.3847/1538-4357/ac2f44}, \href
  {https://ui.adsabs.harvard.edu/abs/2021ApJ...923..214F} {923, 214}

\bibitem[\protect\citeauthoryear{{Faucher-Gigu{\`e}re}}{{Faucher-Gigu{\`e}re}}{2020}]{FaucherGiguere2020}
{Faucher-Gigu{\`e}re} C.-A.,  2020, \mn@doi [\mnras] {10.1093/mnras/staa302},
  \href {https://ui.adsabs.harvard.edu/abs/2020MNRAS.493.1614F} {493, 1614}

\bibitem[\protect\citeauthoryear{{Federrath}, {Schr{\"o}n}, {Banerjee}  \&
  {Klessen}}{{Federrath} et~al.}{2014}]{Federrath2014}
{Federrath} C.,  {Schr{\"o}n} M.,  {Banerjee} R.,   {Klessen} R.~S.,  2014,
  \mn@doi [\apj] {10.1088/0004-637X/790/2/128}, \href
  {https://ui.adsabs.harvard.edu/abs/2014ApJ...790..128F} {790, 128}

\bibitem[\protect\citeauthoryear{Ferland}{Ferland}{2003}]{Ferland2003}
Ferland G.~J.,  2003, \mn@doi [Annual Review of Astronomy and Astrophysics]
  {10.1146/annurev.astro.41.011802.094836}, 41, 517

\bibitem[\protect\citeauthoryear{{Ferland}, {Korista}, {Verner}, {Ferguson},
  {Kingdon}  \& {Verner}}{{Ferland} et~al.}{1998}]{Ferland1998}
{Ferland} G.~J.,  {Korista} K.~T.,  {Verner} D.~A.,  {Ferguson} J.~W.,
  {Kingdon} J.~B.,   {Verner} E.~M.,  1998, \mn@doi [\pasp] {10.1086/316190},
  \href {https://ui.adsabs.harvard.edu/abs/1998PASP..110..761F} {110, 761}

\bibitem[\protect\citeauthoryear{{Fichtner}, {Grassitelli}, {Romano-D{\'\i}az}
  \& {Porciani}}{{Fichtner} et~al.}{2022}]{Fichtner2022}
{Fichtner} Y.~A.,  {Grassitelli} L.,  {Romano-D{\'\i}az} E.,   {Porciani} C.,
  2022, \mn@doi [\mnras] {10.1093/mnras/stac785}, \href
  {https://ui.adsabs.harvard.edu/abs/2022MNRAS.512.4573F} {512, 4573}

\bibitem[\protect\citeauthoryear{Finlay et~al.,}{Finlay
  et~al.}{2012}]{Finlay2012Al26}
Finlay P.,  et~al., 2012, \mn@doi [Phys. Rev. C] {10.1103/PhysRevC.85.055501},
  85, 055501

\bibitem[\protect\citeauthoryear{{Franeck}, {W{\"u}nsch},
  {Mart{\'\i}nez-Gonz{\'a}lez}, {Orlitov{\'a}}, {Boorman}, {Svoboda},
  {Sz{\'e}csi}  \& {Douna}}{{Franeck} et~al.}{2022}]{Franeck2022}
{Franeck} A.,  {W{\"u}nsch} R.,  {Mart{\'\i}nez-Gonz{\'a}lez} S.,
  {Orlitov{\'a}} I.,  {Boorman} P.,  {Svoboda} J.,  {Sz{\'e}csi} D.,   {Douna}
  V.,  2022, \mn@doi [\apj] {10.3847/1538-4357/ac4fc2}, \href
  {https://ui.adsabs.harvard.edu/abs/2022ApJ...927..212F} {927, 212}

\bibitem[\protect\citeauthoryear{{Fuller}, {Piro}  \& {Jermyn}}{{Fuller}
  et~al.}{2019}]{Fuller2019}
{Fuller} J.,  {Piro} A.~L.,   {Jermyn} A.~S.,  2019, \mn@doi [\mnras]
  {10.1093/mnras/stz514}, \href
  {https://ui.adsabs.harvard.edu/abs/2019MNRAS.485.3661F} {485, 3661}

\bibitem[\protect\citeauthoryear{{Gal-Yam} \& {Leonard}}{{Gal-Yam} \&
  {Leonard}}{2009}]{GalYam2009}
{Gal-Yam} A.,  {Leonard} D.~C.,  2009, \mn@doi [\nat] {10.1038/nature07934},
  \href {https://ui.adsabs.harvard.edu/abs/2009Natur.458..865G} {458, 865}

\bibitem[\protect\citeauthoryear{{Garcia}, {Herrero}, {Najarro}, {Camacho}  \&
  {Lorenzo}}{{Garcia} et~al.}{2019}]{Garcia2019}
{Garcia} M.,  {Herrero} A.,  {Najarro} F.,  {Camacho} I.,   {Lorenzo} M.,
  2019, \mn@doi [\mnras] {10.1093/mnras/sty3503}, \href
  {https://ui.adsabs.harvard.edu/abs/2019MNRAS.484..422G} {484, 422}

\bibitem[\protect\citeauthoryear{Geen, Rosdahl, Blaizot, Devriendt  \&
  Slyz}{Geen et~al.}{2015}]{Geen2015}
Geen S.,  Rosdahl J.,  Blaizot J.,  Devriendt J.,   Slyz A.,  2015, \mn@doi
  [Monthly Notices of the Royal Astronomical Society] {10.1093/mnras/stv251},
  448, 3248

\bibitem[\protect\citeauthoryear{{Geen}, {Watson}, {Rosdahl}, {Bieri},
  {Klessen}  \& {Hennebelle}}{{Geen} et~al.}{2018}]{Geen2018}
{Geen} S.,  {Watson} S.~K.,  {Rosdahl} J.,  {Bieri} R.,  {Klessen} R.~S.,
  {Hennebelle} P.,  2018, \mn@doi [\mnras] {10.1093/mnras/sty2439}, \href
  {https://ui.adsabs.harvard.edu/abs/2018MNRAS.481.2548G} {481, 2548}

\bibitem[\protect\citeauthoryear{{Georgy}, {Meynet}, {Ekstr{\"o}m}, {Wade},
  {Petit}, {Keszthelyi}  \& {Hirschi}}{{Georgy} et~al.}{2017}]{Georgy2017}
{Georgy} C.,  {Meynet} G.,  {Ekstr{\"o}m} S.,  {Wade} G.~A.,  {Petit} V.,
  {Keszthelyi} Z.,   {Hirschi} R.,  2017, \mn@doi [\aap]
  {10.1051/0004-6361/201730401}, \href
  {https://ui.adsabs.harvard.edu/abs/2017A&A...599L...5G} {599, L5}

\bibitem[\protect\citeauthoryear{{Gerke}, {Kochanek}  \& {Stanek}}{{Gerke}
  et~al.}{2015}]{Gerke2015}
{Gerke} J.~R.,  {Kochanek} C.~S.,   {Stanek} K.~Z.,  2015, \mn@doi [\mnras]
  {10.1093/mnras/stv776}, \href
  {https://ui.adsabs.harvard.edu/abs/2015MNRAS.450.3289G} {450, 3289}

\bibitem[\protect\citeauthoryear{{Gonzalez-Delgado} et~al.,}{{Gonzalez-Delgado}
  et~al.}{1994}]{GonzalezDelgado1994}
{Gonzalez-Delgado} R.~M.,  et~al., 1994, \mn@doi [\apj] {10.1086/174992}, \href
  {https://ui.adsabs.harvard.edu/abs/1994ApJ...437..239G} {437, 239}

\bibitem[\protect\citeauthoryear{{G{\"o}tberg}, {de Mink}, {Groh}, {Leitherer}
  \& {Norman}}{{G{\"o}tberg} et~al.}{2019}]{Gotberg2019}
{G{\"o}tberg} Y.,  {de Mink} S.~E.,  {Groh} J.~H.,  {Leitherer} C.,   {Norman}
  C.,  2019, \mn@doi [\aap] {10.1051/0004-6361/201834525}, \href
  {https://ui.adsabs.harvard.edu/abs/2019A&A...629A.134G} {629, A134}

\bibitem[\protect\citeauthoryear{{G{\"o}tberg}, {de Mink}, {McQuinn},
  {Zapartas}, {Groh}  \& {Norman}}{{G{\"o}tberg} et~al.}{2020}]{Gotberg2020}
{G{\"o}tberg} Y.,  {de Mink} S.~E.,  {McQuinn} M.,  {Zapartas} E.,  {Groh}
  J.~H.,   {Norman} C.,  2020, \mn@doi [\aap] {10.1051/0004-6361/201936669},
  \href {https://ui.adsabs.harvard.edu/abs/2020A&A...634A.134G} {634, A134}

\bibitem[\protect\citeauthoryear{{Gr{\"a}fener} \& {Vink}}{{Gr{\"a}fener} \&
  {Vink}}{2016}]{Graefener2016}
{Gr{\"a}fener} G.,  {Vink} J.~S.,  2016, \mn@doi [\mnras]
  {10.1093/mnras/stv2283}, \href
  {https://ui.adsabs.harvard.edu/abs/2016MNRAS.455..112G} {455, 112}

\bibitem[\protect\citeauthoryear{{Gr{\"a}fener}, {Owocki}  \&
  {Vink}}{{Gr{\"a}fener} et~al.}{2012}]{Grafener:2012}
{Gr{\"a}fener} G.,  {Owocki} S.~P.,   {Vink} J.~S.,  2012, \mn@doi [\aap]
  {10.1051/0004-6361/201117497}, \href
  {https://ui.adsabs.harvard.edu/abs/2012A&A...538A..40G} {538, A40}

\bibitem[\protect\citeauthoryear{{Grassitelli}, {Langer}, {Mackey},
  {Gr{\"a}fener}, {Grin}, {Sander}  \& {Vink}}{{Grassitelli}
  et~al.}{2021}]{Grassitelli21}
{Grassitelli} L.,  {Langer} N.,  {Mackey} J.,  {Gr{\"a}fener} G.,  {Grin}
  N.~J.,  {Sander} A.~A.~C.,   {Vink} J.~S.,  2021, \mn@doi [\aap]
  {10.1051/0004-6361/202038298}, \href
  {https://ui.adsabs.harvard.edu/abs/2021A&A...647A..99G} {647, A99}

\bibitem[\protect\citeauthoryear{Groenewegen, Lamers  \& Pauldrach}{Groenewegen
  et~al.}{1989}]{Groenewegen1989}
Groenewegen M.,  Lamers H.,   Pauldrach A.,  1989, Astronomy \& Astrophysics,
  221, 78

\bibitem[\protect\citeauthoryear{{Groh}, {Meynet}  \& {Ekstr{\"o}m}}{{Groh}
  et~al.}{2013a}]{Groh2013b}
{Groh} J.~H.,  {Meynet} G.,   {Ekstr{\"o}m} S.,  2013a, \mn@doi [\aap]
  {10.1051/0004-6361/201220741}, \href
  {https://ui.adsabs.harvard.edu/abs/2013A&A...550L...7G} {550, L7}

\bibitem[\protect\citeauthoryear{{Groh}, {Meynet}, {Georgy}  \&
  {Ekstr{\"o}m}}{{Groh} et~al.}{2013b}]{Groh2013a}
{Groh} J.~H.,  {Meynet} G.,  {Georgy} C.,   {Ekstr{\"o}m} S.,  2013b, \mn@doi
  [\aap] {10.1051/0004-6361/201321906}, \href
  {https://ui.adsabs.harvard.edu/abs/2013A&A...558A.131G} {558, A131}

\bibitem[\protect\citeauthoryear{{Grudi{\'c}}, {Guszejnov}, {Offner}, {Rosen},
  {Raju}, {Faucher-Gigu{\`e}re}  \& {Hopkins}}{{Grudi{\'c}}
  et~al.}{2022}]{Grudic2022}
{Grudi{\'c}} M.~Y.,  {Guszejnov} D.,  {Offner} S. S.~R.,  {Rosen} A.~L.,
  {Raju} A.~N.,  {Faucher-Gigu{\`e}re} C.-A.,   {Hopkins} P.~F.,  2022, \mn@doi
  [\mnras] {10.1093/mnras/stac526}, \href
  {https://ui.adsabs.harvard.edu/abs/2022MNRAS.512..216G} {512, 216}

\bibitem[\protect\citeauthoryear{Guedel, Briggs, Montmerle, Audard, Rebull  \&
  Skinner}{Guedel et~al.}{2008}]{Guedel2008}
Guedel M.,  Briggs K.~R.,  Montmerle T.,  Audard M.,  Rebull L.,   Skinner
  S.~L.,  2008, \mn@doi [Science] {10.1126/science.1149926}, 319, 309

\bibitem[\protect\citeauthoryear{{Guidi}, {Scannapieco}  \& {Walcher}}{{Guidi}
  et~al.}{2015}]{Guidi2015}
{Guidi} G.,  {Scannapieco} C.,   {Walcher} C.~J.,  2015, \mn@doi [\mnras]
  {10.1093/mnras/stv2050}, \href
  {https://ui.adsabs.harvard.edu/abs/2015MNRAS.454.2381G} {454, 2381}

\bibitem[\protect\citeauthoryear{{Gutcke}, {Pakmor}, {Naab}  \&
  {Springel}}{{Gutcke} et~al.}{2021}]{Gutcke2021}
{Gutcke} T.~A.,  {Pakmor} R.,  {Naab} T.,   {Springel} V.,  2021, \mn@doi
  [\mnras] {10.1093/mnras/staa3875}, \href
  {https://ui.adsabs.harvard.edu/abs/2021MNRAS.501.5597G} {501, 5597}

\bibitem[\protect\citeauthoryear{{Haworth}, {Glover}, {Koepferl}, {Bisbas}  \&
  {Dale}}{{Haworth} et~al.}{2018}]{Haworth2018}
{Haworth} T.~J.,  {Glover} S. C.~O.,  {Koepferl} C.~M.,  {Bisbas} T.~G.,
  {Dale} J.~E.,  2018, \mn@doi [\nar] {10.1016/j.newar.2018.06.001}, \href
  {https://ui.adsabs.harvard.edu/abs/2018NewAR..82....1H} {82, 1}

\bibitem[\protect\citeauthoryear{{Heger}, {Woosley}  \& {Spruit}}{{Heger}
  et~al.}{2005}]{Heger2005}
{Heger} A.,  {Woosley} S.~E.,   {Spruit} H.~C.,  2005, \mn@doi [\apj]
  {10.1086/429868}, \href
  {https://ui.adsabs.harvard.edu/abs/2005ApJ...626..350H} {626, 350}

\bibitem[\protect\citeauthoryear{{Henry} \& {Worthey}}{{Henry} \&
  {Worthey}}{1999}]{Henry1999}
{Henry} R.~B.~C.,  {Worthey} G.,  1999, \mn@doi [\pasp] {10.1086/316403}, \href
  {https://ui.adsabs.harvard.edu/abs/1999PASP..111..919H} {111, 919}

\bibitem[\protect\citeauthoryear{{Higgins} \& {Vink}}{{Higgins} \&
  {Vink}}{2020}]{Higgins2020}
{Higgins} E.~R.,  {Vink} J.~S.,  2020, \mn@doi [\aap]
  {10.1051/0004-6361/201937374}, \href
  {https://ui.adsabs.harvard.edu/abs/2020A&A...635A.175H} {635, A175}

\bibitem[\protect\citeauthoryear{{Hirschmann} et~al.,}{{Hirschmann}
  et~al.}{2013}]{Hirschmann2013}
{Hirschmann} M.,  et~al., 2013, \mn@doi [\mnras] {10.1093/mnras/stt1770}, \href
  {https://ui.adsabs.harvard.edu/abs/2013MNRAS.436.2929H} {436, 2929}

\bibitem[\protect\citeauthoryear{{Horiuchi}, {Nakamura}, {Takiwaki}, {Kotake}
  \& {Tanaka}}{{Horiuchi} et~al.}{2014}]{Horiuchi2014}
{Horiuchi} S.,  {Nakamura} K.,  {Takiwaki} T.,  {Kotake} K.,   {Tanaka} M.,
  2014, \mn@doi [\mnras] {10.1093/mnrasl/slu146}, \href
  {https://ui.adsabs.harvard.edu/abs/2014MNRAS.445L..99H} {445, L99}

\bibitem[\protect\citeauthoryear{{Humphreys} \& {Davidson}}{{Humphreys} \&
  {Davidson}}{1994}]{Humphreys1994}
{Humphreys} R.~M.,  {Davidson} K.,  1994, \mn@doi [\pasp] {10.1086/133478},
  \href {https://ui.adsabs.harvard.edu/abs/1994PASP..106.1025H} {106, 1025}

\bibitem[\protect\citeauthoryear{{Izotov}, {Guseva}, {Lipovetskii}, {Kniazev}
  \& {Stepanian}}{{Izotov} et~al.}{1990}]{Izotov1990}
{Izotov} I.~I.,  {Guseva} N.~G.,  {Lipovetskii} V.~A.,  {Kniazev} A.~I.,
  {Stepanian} J.~A.,  1990, \mn@doi [\nat] {10.1038/343238a0}, \href
  {https://ui.adsabs.harvard.edu/abs/1990Natur.343..238I} {343, 238}

\bibitem[\protect\citeauthoryear{{Izotov}, {Schaerer}, {Thuan}, {Worseck},
  {Guseva}, {Orlitov{\'a}}  \& {Verhamme}}{{Izotov} et~al.}{2016}]{Izotov2016}
{Izotov} Y.~I.,  {Schaerer} D.,  {Thuan} T.~X.,  {Worseck} G.,  {Guseva} N.~G.,
   {Orlitov{\'a}} I.,   {Verhamme} A.,  2016, \mn@doi [\mnras]
  {10.1093/mnras/stw1205}, \href
  {https://ui.adsabs.harvard.edu/abs/2016MNRAS.461.3683I} {461, 3683}

\bibitem[\protect\citeauthoryear{{Jiang}, {Cantiello}, {Bildsten}, {Quataert}
  \& {Blaes}}{{Jiang} et~al.}{2015}]{Jiang2015}
{Jiang} Y.-F.,  {Cantiello} M.,  {Bildsten} L.,  {Quataert} E.,   {Blaes} O.,
  2015, \mn@doi [\apj] {10.1088/0004-637X/813/1/74}, \href
  {https://ui.adsabs.harvard.edu/abs/2015ApJ...813...74J} {813, 74}

\bibitem[\protect\citeauthoryear{{Jones}, {Swinbank}, {Ellis}, {Richard}  \&
  {Stark}}{{Jones} et~al.}{2010}]{Jones2010}
{Jones} T.~A.,  {Swinbank} A.~M.,  {Ellis} R.~S.,  {Richard} J.,   {Stark}
  D.~P.,  2010, \mn@doi [\mnras] {10.1111/j.1365-2966.2010.16378.x}, \href
  {https://ui.adsabs.harvard.edu/abs/2010MNRAS.404.1247J} {404, 1247}

\bibitem[\protect\citeauthoryear{{Kaiser}, {Hirschi}, {Arnett}, {Georgy},
  {Scott}  \& {Cristini}}{{Kaiser} et~al.}{2020}]{Kaiser:2020}
{Kaiser} E.~A.,  {Hirschi} R.,  {Arnett} W.~D.,  {Georgy} C.,  {Scott} L.
  J.~A.,   {Cristini} A.,  2020, \mn@doi [\mnras] {10.1093/mnras/staa1595},
  \href {https://ui.adsabs.harvard.edu/abs/2020MNRAS.496.1967K} {496, 1967}

\bibitem[\protect\citeauthoryear{{Kasen}, {Metzger}, {Barnes}, {Quataert}  \&
  {Ramirez-Ruiz}}{{Kasen} et~al.}{2017a}]{Kasen17}
{Kasen} D.,  {Metzger} B.,  {Barnes} J.,  {Quataert} E.,   {Ramirez-Ruiz} E.,
  2017a, \mn@doi [\nat] {10.1038/nature24453}, \href
  {https://ui.adsabs.harvard.edu/abs/2017Natur.551...80K} {551, 80}

\bibitem[\protect\citeauthoryear{{Kasen}, {Metzger}, {Barnes}, {Quataert}  \&
  {Ramirez-Ruiz}}{{Kasen} et~al.}{2017b}]{Kasen2017}
{Kasen} D.,  {Metzger} B.,  {Barnes} J.,  {Quataert} E.,   {Ramirez-Ruiz} E.,
  2017b, \mn@doi [\nat] {10.1038/nature24453}, \href
  {https://ui.adsabs.harvard.edu/abs/2017Natur.551...80K} {551, 80}

\bibitem[\protect\citeauthoryear{{Katz}}{{Katz}}{1992}]{Katz1992}
{Katz} N.,  1992, \mn@doi [\apj] {10.1086/171366}, \href
  {https://ui.adsabs.harvard.edu/abs/1992ApJ...391..502K} {391, 502}

\bibitem[\protect\citeauthoryear{{Kawata}}{{Kawata}}{2001}]{Kawata2001}
{Kawata} D.,  2001, \mn@doi [\apj] {10.1086/322309}, \href
  {https://ui.adsabs.harvard.edu/abs/2001ApJ...558..598K} {558, 598}

\bibitem[\protect\citeauthoryear{{Kee}, {Sundqvist}, {Decin}, {de Koter}  \&
  {Sana}}{{Kee} et~al.}{2021}]{Kee2021}
{Kee} N.~D.,  {Sundqvist} J.~O.,  {Decin} L.,  {de Koter} A.,   {Sana} H.,
  2021, \mn@doi [\aap] {10.1051/0004-6361/202039224}, \href
  {https://ui.adsabs.harvard.edu/abs/2021A&A...646A.180K} {646, A180}

\bibitem[\protect\citeauthoryear{{Keller} \& {Kruijssen}}{{Keller} \&
  {Kruijssen}}{2022}]{Keller2022}
{Keller} B.~W.,  {Kruijssen} J.~M.~D.,  2022, \mn@doi [\mnras]
  {10.1093/mnras/stac511}, \href
  {https://ui.adsabs.harvard.edu/abs/2022MNRAS.512..199K} {512, 199}

\bibitem[\protect\citeauthoryear{{Kennicutt}, {Lee}, {Funes}, {J.}, {Sakai}  \&
  {Akiyama}}{{Kennicutt} et~al.}{2008}]{Kennicutt2008}
{Kennicutt} Robert~C. J.,  {Lee} J.~C.,  {Funes} J.~G.,  {J.} S.,  {Sakai} S.,
   {Akiyama} S.,  2008, \mn@doi [\apjs] {10.1086/590058}, \href
  {https://ui.adsabs.harvard.edu/abs/2008ApJS..178..247K} {178, 247}

\bibitem[\protect\citeauthoryear{{Keszthelyi}, {Wade}  \& {Petit}}{{Keszthelyi}
  et~al.}{2017a}]{Keszthelyi2017a}
{Keszthelyi} Z.,  {Wade} G.~A.,   {Petit} V.,  2017a, in {Eldridge} J.~J.,
  {Bray} J.~C.,  {McClelland} L.~A.~S.,   {Xiao} L.,  eds, ~ Vol. 329, The
  Lives and Death-Throes of Massive Stars. pp 250--254 (\mn@eprint {arXiv}
  {1702.04460}), \mn@doi{10.1017/S1743921317002745}

\bibitem[\protect\citeauthoryear{{Keszthelyi}, {Puls}  \& {Wade}}{{Keszthelyi}
  et~al.}{2017b}]{Keszthelyi2017b}
{Keszthelyi} Z.,  {Puls} J.,   {Wade} G.~A.,  2017b, \mn@doi [\aap]
  {10.1051/0004-6361/201629468}, \href
  {https://ui.adsabs.harvard.edu/abs/2017A&A...598A...4K} {598, A4}

\bibitem[\protect\citeauthoryear{{Keszthelyi}, {Meynet}, {Georgy}, {Wade},
  {Petit}  \& {David-Uraz}}{{Keszthelyi} et~al.}{2019}]{Keszthelyi2019}
{Keszthelyi} Z.,  {Meynet} G.,  {Georgy} C.,  {Wade} G.~A.,  {Petit} V.,
  {David-Uraz} A.,  2019, \mn@doi [\mnras] {10.1093/mnras/stz772}, \href
  {https://ui.adsabs.harvard.edu/abs/2019MNRAS.485.5843K} {485, 5843}

\bibitem[\protect\citeauthoryear{{Keszthelyi} et~al.,}{{Keszthelyi}
  et~al.}{2020}]{Keszthelyi2020}
{Keszthelyi} Z.,  et~al., 2020, \mn@doi [\mnras] {10.1093/mnras/staa237}, \href
  {https://ui.adsabs.harvard.edu/abs/2020MNRAS.493..518K} {493, 518}

\bibitem[\protect\citeauthoryear{{Keszthelyi}, {Meynet}, {Martins}, {de Koter}
  \& {David-Uraz}}{{Keszthelyi} et~al.}{2021}]{Keszthelyi2021}
{Keszthelyi} Z.,  {Meynet} G.,  {Martins} F.,  {de Koter} A.,   {David-Uraz}
  A.,  2021, \mn@doi [\mnras] {10.1093/mnras/stab893}, \href
  {https://ui.adsabs.harvard.edu/abs/2021MNRAS.504.2474K} {504, 2474}

\bibitem[\protect\citeauthoryear{{Keszthelyi} et~al.,}{{Keszthelyi}
  et~al.}{2022}]{Keszthelyi2022}
{Keszthelyi} Z.,  et~al., 2022, \mn@doi [\mnras] {10.1093/mnras/stac2598},
  \href {https://ui.adsabs.harvard.edu/abs/2022MNRAS.tmp.2418K}
  {10.1093/mnras/stac2598}

\bibitem[\protect\citeauthoryear{{Kobayashi}, {Karakas}  \&
  {Lugaro}}{{Kobayashi} et~al.}{2020}]{kobayashi2020}
{Kobayashi} C.,  {Karakas} A.~I.,   {Lugaro} M.,  2020, \mn@doi [\apj]
  {10.3847/1538-4357/abae65}, \href
  {https://ui.adsabs.harvard.edu/abs/2020ApJ...900..179K} {900, 179}

\bibitem[\protect\citeauthoryear{{Kotak} \& {Vink}}{{Kotak} \&
  {Vink}}{2006}]{Kotak2006}
{Kotak} R.,  {Vink} J.~S.,  2006, \mn@doi [\aap] {10.1051/0004-6361:20065800},
  \href {https://ui.adsabs.harvard.edu/abs/2006A&A...460L...5K} {460, L5}

\bibitem[\protect\citeauthoryear{{Krumholz} \& {Burkhart}}{{Krumholz} \&
  {Burkhart}}{2016}]{Krumholz2016}
{Krumholz} M.~R.,  {Burkhart} B.,  2016, \mn@doi [\mnras]
  {10.1093/mnras/stw434}, \href
  {https://ui.adsabs.harvard.edu/abs/2016MNRAS.458.1671K} {458, 1671}

\bibitem[\protect\citeauthoryear{{Kudritzki} \& {Puls}}{{Kudritzki} \&
  {Puls}}{2000}]{Kudritzki2000}
{Kudritzki} R.-P.,  {Puls} J.,  2000, \araa, 38, 613

\bibitem[\protect\citeauthoryear{{Kuiper} \& {Hosokawa}}{{Kuiper} \&
  {Hosokawa}}{2018}]{Kuiper2018}
{Kuiper} R.,  {Hosokawa} T.,  2018, \mn@doi [\aap]
  {10.1051/0004-6361/201832638}, \href
  {https://ui.adsabs.harvard.edu/abs/2018A&A...616A.101K} {616, A101}

\bibitem[\protect\citeauthoryear{Lancaster, Ostriker, Kim  \& Kim}{Lancaster
  et~al.}{2021}]{Lancaster2021a}
Lancaster L.,  Ostriker E.~C.,  Kim J.-G.,   Kim C.-G.,  2021, \mn@doi [The
  Astrophysical Journal] {10.3847/1538-4357/abf8ab}, 914, 89

\bibitem[\protect\citeauthoryear{{Langer}}{{Langer}}{1997}]{Langer1997}
{Langer} N.,  1997, in {Nota} A.,  {Lamers} H.,  eds,  Astronomical Society of
  the Pacific Conference Series Vol. 120, Luminous Blue Variables: Massive
  Stars in Transition. p.~83

\bibitem[\protect\citeauthoryear{{Langer}}{{Langer}}{1998}]{Langer1998}
{Langer} N.,  1998, \aap, \href
  {https://ui.adsabs.harvard.edu/abs/1998A&A...329..551L} {329, 551}

\bibitem[\protect\citeauthoryear{{Laplace}, {Justham}, {Renzo}, {G{\"o}tberg},
  {Farmer}, {Vartanyan}  \& {de Mink}}{{Laplace} et~al.}{2021}]{Laplace2021}
{Laplace} E.,  {Justham} S.,  {Renzo} M.,  {G{\"o}tberg} Y.,  {Farmer} R.,
  {Vartanyan} D.,   {de Mink} S.~E.,  2021, \mn@doi [\aap]
  {10.1051/0004-6361/202140506}, \href
  {https://ui.adsabs.harvard.edu/abs/2021A&A...656A..58L} {656, A58}

\bibitem[\protect\citeauthoryear{{Larson}}{{Larson}}{1974}]{Larson1974}
{Larson} R.~B.,  1974, \mn@doi [\mnras] {10.1093/mnras/169.2.229}, \href
  {https://ui.adsabs.harvard.edu/abs/1974MNRAS.169..229L} {169, 229}

\bibitem[\protect\citeauthoryear{{Leitherer} et~al.,}{{Leitherer}
  et~al.}{1999}]{Leitherer1999}
{Leitherer} C.,  et~al., 1999, \mn@doi [\apjs] {10.1086/313233}, \href
  {https://ui.adsabs.harvard.edu/abs/1999ApJS..123....3L} {123, 3}

\bibitem[\protect\citeauthoryear{{Lemasle} et~al.,}{{Lemasle}
  et~al.}{2018}]{Lemasle2018}
{Lemasle} B.,  et~al., 2018, \mn@doi [\aap] {10.1051/0004-6361/201834050},
  \href {https://ui.adsabs.harvard.edu/abs/2018A&A...618A.160L} {618, A160}

\bibitem[\protect\citeauthoryear{{Lequeux}, {Peimbert}, {Rayo}, {Serrano}  \&
  {Torres-Peimbert}}{{Lequeux} et~al.}{1979}]{Lequeux1979}
{Lequeux} J.,  {Peimbert} M.,  {Rayo} J.~F.,  {Serrano} A.,   {Torres-Peimbert}
  S.,  1979, \aap, \href
  {https://ui.adsabs.harvard.edu/abs/1979A&A....80..155L} {80, 155}

\bibitem[\protect\citeauthoryear{{Livermore} et~al.,}{{Livermore}
  et~al.}{2015}]{Livermore2015}
{Livermore} R.~C.,  et~al., 2015, \mn@doi [\mnras] {10.1093/mnras/stv686},
  \href {https://ui.adsabs.harvard.edu/abs/2015MNRAS.450.1812L} {450, 1812}

\bibitem[\protect\citeauthoryear{{Lopez}, {Krumholz}, {Bolatto}, {Prochaska},
  {Ramirez-Ruiz}  \& {Castro}}{{Lopez} et~al.}{2014}]{lopez14}
{Lopez} L.~A.,  {Krumholz} M.~R.,  {Bolatto} A.~D.,  {Prochaska} J.~X.,
  {Ramirez-Ruiz} E.,   {Castro} D.,  2014, \mn@doi [\apj]
  {10.1088/0004-637X/795/2/121}, \href
  {http://adsabs.harvard.edu/abs/2014ApJ...795..121L} {795, 121}

\bibitem[\protect\citeauthoryear{{Lorenzo}, {Garcia}, {Najarro}, {Herrero},
  {Cervi{\~n}o}  \& {Castro}}{{Lorenzo} et~al.}{2022}]{Lorenzo2022}
{Lorenzo} M.,  {Garcia} M.,  {Najarro} F.,  {Herrero} A.,  {Cervi{\~n}o} M.,
  {Castro} N.,  2022, \mn@doi [\mnras] {10.1093/mnras/stac2050}, \href
  {https://ui.adsabs.harvard.edu/abs/2022MNRAS.516.4164L} {516, 4164}

\bibitem[\protect\citeauthoryear{{Lovegrove} \& {Woosley}}{{Lovegrove} \&
  {Woosley}}{2013}]{Lovegrove2013}
{Lovegrove} E.,  {Woosley} S.~E.,  2013, \mn@doi [\apj]
  {10.1088/0004-637X/769/2/109}, \href
  {https://ui.adsabs.harvard.edu/abs/2013ApJ...769..109L} {769, 109}

\bibitem[\protect\citeauthoryear{{Lucas}, {Bonnell}  \& {Dale}}{{Lucas}
  et~al.}{2020}]{lucas2020}
{Lucas} W.~E.,  {Bonnell} I.~A.,   {Dale} J.~E.,  2020, \mn@doi [\mnras]
  {10.1093/mnras/staa451}, \href
  {https://ui.adsabs.harvard.edu/abs/2020MNRAS.493.4700L} {493, 4700}

\bibitem[\protect\citeauthoryear{{Mackey}, {Castro}, {Fossati}  \&
  {Langer}}{{Mackey} et~al.}{2015}]{Mackey15}
{Mackey} J.,  {Castro} N.,  {Fossati} L.,   {Langer} N.,  2015, \mn@doi [\aap]
  {10.1051/0004-6361/201526159}, \href
  {https://ui.adsabs.harvard.edu/abs/2015A&A...582A..24M} {582, A24}

\bibitem[\protect\citeauthoryear{{Madau} \& {Dickinson}}{{Madau} \&
  {Dickinson}}{2014}]{MadauDickinson2014}
{Madau} P.,  {Dickinson} M.,  2014, \mn@doi [\araa]
  {10.1146/annurev-astro-081811-125615}, \href
  {https://ui.adsabs.harvard.edu/abs/2014ARA&A..52..415M} {52, 415}

\bibitem[\protect\citeauthoryear{{Maeder}}{{Maeder}}{1987}]{Maeder87}
{Maeder} A.,  1987, \aap, \href
  {https://ui.adsabs.harvard.edu/abs/1987A&A...178..159M} {178, 159}

\bibitem[\protect\citeauthoryear{{Maeder}}{{Maeder}}{2009}]{Maeder2009}
{Maeder} A.,  2009, {Physics, Formation and Evolution of Rotating Stars}.
Springer Berlin Heidelberg, \mn@doi{10.1007/978-3-540-76949-1}

\bibitem[\protect\citeauthoryear{{Maeder} \& {Meynet}}{{Maeder} \&
  {Meynet}}{2000}]{Maeder2000}
{Maeder} A.,  {Meynet} G.,  2000, \mn@doi [\araa]
  {10.1146/annurev.astro.38.1.143}, \href
  {https://ui.adsabs.harvard.edu/abs/2000ARA&A..38..143M} {38, 143}

\bibitem[\protect\citeauthoryear{{Maeder} \& {Meynet}}{{Maeder} \&
  {Meynet}}{2003}]{Maeder2003}
{Maeder} A.,  {Meynet} G.,  2003, \mn@doi [\aap] {10.1051/0004-6361:20031491},
  \href {https://ui.adsabs.harvard.edu/abs/2003A&A...411..543M} {411, 543}

\bibitem[\protect\citeauthoryear{{Maeder} \& {Meynet}}{{Maeder} \&
  {Meynet}}{2004}]{Maeder2004}
{Maeder} A.,  {Meynet} G.,  2004, \mn@doi [\aap] {10.1051/0004-6361:20034583},
  \href {https://ui.adsabs.harvard.edu/abs/2004A&A...422..225M} {422, 225}

\bibitem[\protect\citeauthoryear{{Maeder} \& {Meynet}}{{Maeder} \&
  {Meynet}}{2005}]{Maeder2005}
{Maeder} A.,  {Meynet} G.,  2005, \mn@doi [\aap] {10.1051/0004-6361:20053261},
  \href {https://ui.adsabs.harvard.edu/abs/2005A&A...440.1041M} {440, 1041}

\bibitem[\protect\citeauthoryear{{Mart{\'\i}nez-Gonz{\'a}lez}, {W{\"u}nsch},
  {Tenorio-Tagle}, {Silich}, {Sz{\'e}csi}  \&
  {Palou{\v{s}}}}{{Mart{\'\i}nez-Gonz{\'a}lez}
  et~al.}{2022}]{MartinezGonzalez2022}
{Mart{\'\i}nez-Gonz{\'a}lez} S.,  {W{\"u}nsch} R.,  {Tenorio-Tagle} G.,
  {Silich} S.,  {Sz{\'e}csi} D.,   {Palou{\v{s}}} J.,  2022, \mn@doi [\apj]
  {10.3847/1538-4357/ac77fe}, \href
  {https://ui.adsabs.harvard.edu/abs/2022ApJ...934...51M} {934, 51}

\bibitem[\protect\citeauthoryear{{Martins}, {Schaerer}, {Hillier}, {Meynadier},
  {Heydari-Malayeri}  \& {Walborn}}{{Martins} et~al.}{2005}]{Martins2005}
{Martins} F.,  {Schaerer} D.,  {Hillier} D.~J.,  {Meynadier} F.,
  {Heydari-Malayeri} M.,   {Walborn} N.~R.,  2005, \aap, 441, 735

\bibitem[\protect\citeauthoryear{{Mathews} \& {Baker}}{{Mathews} \&
  {Baker}}{1971}]{Mathews1971}
{Mathews} W.~G.,  {Baker} J.~C.,  1971, \mn@doi [\apj] {10.1086/151208}, \href
  {https://ui.adsabs.harvard.edu/abs/1971ApJ...170..241M} {170, 241}

\bibitem[\protect\citeauthoryear{{McCray} \& {Snow}}{{McCray} \&
  {Snow}}{1979}]{McCray1979}
{McCray} R.,  {Snow} T.~P. J.,  1979, \mn@doi [\araa]
  {10.1146/annurev.aa.17.090179.001241}, \href
  {https://ui.adsabs.harvard.edu/abs/1979ARA&A..17..213M} {17, 213}

\bibitem[\protect\citeauthoryear{{McDonald}, {Davies}  \& {Beasor}}{{McDonald}
  et~al.}{2022}]{McDonald22}
{McDonald} S. L.~E.,  {Davies} B.,   {Beasor} E.~R.,  2022, \mn@doi [\mnras]
  {10.1093/mnras/stab3453}, \href
  {https://ui.adsabs.harvard.edu/abs/2022MNRAS.510.3132M} {510, 3132}

\bibitem[\protect\citeauthoryear{{McKee} \& {Ostriker}}{{McKee} \&
  {Ostriker}}{1977}]{McKee1977}
{McKee} C.~F.,  {Ostriker} J.~P.,  1977, \mn@doi [\apj] {10.1086/155667}, \href
  {https://ui.adsabs.harvard.edu/abs/1977ApJ...218..148M} {218, 148}

\bibitem[\protect\citeauthoryear{{McLeod}, {Dale}, {Evans}, {Ginsburg},
  {Kruijssen}, {Pellegrini}, {Ramsay}  \& {Testi}}{{McLeod}
  et~al.}{2019}]{mcleod18}
{McLeod} A.~F.,  {Dale} J.~E.,  {Evans} C.~J.,  {Ginsburg} A.,  {Kruijssen}
  J.~M.~D.,  {Pellegrini} E.~W.,  {Ramsay} S.~K.,   {Testi} L.,  2019, \mn@doi
  [\mnras] {10.1093/mnras/sty2696}, \href
  {https://ui.adsabs.harvard.edu/abs/2019MNRAS.486.5263M} {486, 5263}

\bibitem[\protect\citeauthoryear{{McLeod} et~al.,}{{McLeod}
  et~al.}{2020}]{mcleod20}
{McLeod} A.~F.,  et~al., 2020, \mn@doi [\apj] {10.3847/1538-4357/ab6d63}, \href
  {https://ui.adsabs.harvard.edu/abs/2020ApJ...891...25M} {891, 25}

\bibitem[\protect\citeauthoryear{{McLeod} et~al.,}{{McLeod}
  et~al.}{2021}]{mcleod21}
{McLeod} A.~F.,  et~al., 2021, \mn@doi [\mnras] {10.1093/mnras/stab2726}, \href
  {https://ui.adsabs.harvard.edu/abs/2021MNRAS.508.5425M} {508, 5425}

\bibitem[\protect\citeauthoryear{{Meynet}, {Eggenberger}  \& {Maeder}}{{Meynet}
  et~al.}{2011}]{Meynet2011}
{Meynet} G.,  {Eggenberger} P.,   {Maeder} A.,  2011, \aap, 525, L11

\bibitem[\protect\citeauthoryear{{Meynet} et~al.,}{{Meynet}
  et~al.}{2015}]{Meynet2015}
{Meynet} G.,  et~al., 2015, \mn@doi [\aap] {10.1051/0004-6361/201424671}, \href
  {https://ui.adsabs.harvard.edu/abs/2015A&A...575A..60M} {575, A60}

\bibitem[\protect\citeauthoryear{{Micheva}, {Oey}, {Jaskot}  \&
  {James}}{{Micheva} et~al.}{2017}]{Micheva2017}
{Micheva} G.,  {Oey} M.~S.,  {Jaskot} A.~E.,   {James} B.~L.,  2017, \mn@doi
  [\apj] {10.3847/1538-4357/aa830b}, \href
  {https://ui.adsabs.harvard.edu/abs/2017ApJ...845..165M} {845, 165}

\bibitem[\protect\citeauthoryear{{Moe} \& {Di Stefano}}{{Moe} \& {Di
  Stefano}}{2017}]{Moe:2017}
{Moe} M.,  {Di Stefano} R.,  2017, \mn@doi [\apjs] {10.3847/1538-4365/aa6fb6},
  \href {https://ui.adsabs.harvard.edu/abs/2017ApJS..230...15M} {230, 15}

\bibitem[\protect\citeauthoryear{{Montarg{\`e}s} et~al.,}{{Montarg{\`e}s}
  et~al.}{2021}]{Montarges2021}
{Montarg{\`e}s} M.,  et~al., 2021, \nat, 594, 365

\bibitem[\protect\citeauthoryear{{M{\"u}ller} et~al.,}{{M{\"u}ller}
  et~al.}{2019}]{Mueller2019}
{M{\"u}ller} B.,  et~al., 2019, \mn@doi [\mnras] {10.1093/mnras/stz216}, \href
  {https://ui.adsabs.harvard.edu/abs/2019MNRAS.484.3307M} {484, 3307}

\bibitem[\protect\citeauthoryear{{Naab} \& {Ostriker}}{{Naab} \&
  {Ostriker}}{2017}]{Naab2017}
{Naab} T.,  {Ostriker} J.~P.,  2017, \mn@doi [\araa]
  {10.1146/annurev-astro-081913-040019}, \href
  {https://ui.adsabs.harvard.edu/abs/2017ARA&A..55...59N} {55, 59}

\bibitem[\protect\citeauthoryear{{Nogueras-Lara} et~al.,}{{Nogueras-Lara}
  et~al.}{2018}]{NoguerasLara2018}
{Nogueras-Lara} F.,  et~al., 2018, \mn@doi [\aap]
  {10.1051/0004-6361/201833518}, \href
  {https://ui.adsabs.harvard.edu/abs/2018A&A...620A..83N} {620, A83}

\bibitem[\protect\citeauthoryear{{Olivier}, {Berg}, {Chisholm}, {Erb}, {Pogge}
  \& {Skillman}}{{Olivier} et~al.}{2021a}]{Olivier2021b}
{Olivier} G.~M.,  {Berg} D.~A.,  {Chisholm} J.,  {Erb} D.~K.,  {Pogge} R.~W.,
  {Skillman} E.~D.,  2021a, arXiv e-prints, \href
  {https://ui.adsabs.harvard.edu/abs/2021arXiv210906725O} {p. arXiv:2109.06725}

\bibitem[\protect\citeauthoryear{{Olivier}, {Lopez}, {Rosen}, {Nayak},
  {Reiter}, {Krumholz}  \& {Bolatto}}{{Olivier} et~al.}{2021b}]{Olivier21a}
{Olivier} G.~M.,  {Lopez} L.~A.,  {Rosen} A.~L.,  {Nayak} O.,  {Reiter} M.,
  {Krumholz} M.~R.,   {Bolatto} A.~D.,  2021b, \mn@doi [\apj]
  {10.3847/1538-4357/abd24a}, \href
  {https://ui.adsabs.harvard.edu/abs/2021ApJ...908...68O} {908, 68}

\bibitem[\protect\citeauthoryear{{Oskinova} \& {Schaerer}}{{Oskinova} \&
  {Schaerer}}{2022}]{oskinova2022}
{Oskinova} L.,  {Schaerer} D.,  2022, arXiv e-prints, \href
  {https://ui.adsabs.harvard.edu/abs/2022arXiv220304987O} {p. arXiv:2203.04987}

\bibitem[\protect\citeauthoryear{{Owocki}}{{Owocki}}{2004}]{Owocki2004}
{Owocki} S.~P.,  2004, in {Maeder} A.,  {Eenens} P.,  eds,  IAU Symposium Vol.
  215, Stellar Rotation. p.~515

\bibitem[\protect\citeauthoryear{{Padoan}, {Pan}, {Haugb{\o}lle}  \&
  {Nordlund}}{{Padoan} et~al.}{2016}]{Padoan2016}
{Padoan} P.,  {Pan} L.,  {Haugb{\o}lle} T.,   {Nordlund} {\r{A}}.,  2016,
  \mn@doi [\apj] {10.3847/0004-637X/822/1/11}, \href
  {https://ui.adsabs.harvard.edu/abs/2016ApJ...822...11P} {822, 11}

\bibitem[\protect\citeauthoryear{{Paxton} et~al.,}{{Paxton}
  et~al.}{2013}]{Paxton2013}
{Paxton} B.,  et~al., 2013, \mn@doi [\apjs] {10.1088/0067-0049/208/1/4}, \href
  {http://adsabs.harvard.edu/abs/2013ApJS..208....4P} {208, 4}

\bibitem[\protect\citeauthoryear{{Pedersen} et~al.,}{{Pedersen}
  et~al.}{2021}]{Pedersen2021}
{Pedersen} M.~G.,  et~al., 2021, \mn@doi [Nature Astronomy]
  {10.1038/s41550-021-01351-x}, \href
  {https://ui.adsabs.harvard.edu/abs/2021NatAs...5..715P} {5, 715}

\bibitem[\protect\citeauthoryear{{Peimbert}, {Torres-Peimbert}  \&
  {Rayo}}{{Peimbert} et~al.}{1978}]{Peimbert78}
{Peimbert} M.,  {Torres-Peimbert} S.,   {Rayo} J.~F.,  1978, \mn@doi [\apj]
  {10.1086/155933}, \href
  {https://ui.adsabs.harvard.edu/abs/1978ApJ...220..516P} {220, 516}

\bibitem[\protect\citeauthoryear{{Pellegrini}, {Rahner}, {Reissl}, {Glover},
  {Klessen}, {Rousseau-Nepton}  \& {Herrera-Camus}}{{Pellegrini}
  et~al.}{2020}]{Pellegrini2020}
{Pellegrini} E.~W.,  {Rahner} D.,  {Reissl} S.,  {Glover} S.~C.~O.,  {Klessen}
  R.~S.,  {Rousseau-Nepton} L.,   {Herrera-Camus} R.,  2020, \mn@doi [\mnras]
  {10.1093/mnras/staa1473}, \href
  {https://ui.adsabs.harvard.edu/abs/2020MNRAS.496..339P} {496, 339}

\bibitem[\protect\citeauthoryear{{Petit} et~al.,}{{Petit}
  et~al.}{2017}]{Petit2017}
{Petit} V.,  et~al., 2017, \mn@doi [\mnras] {10.1093/mnras/stw3126}, \href
  {https://ui.adsabs.harvard.edu/abs/2017MNRAS.466.1052P} {466, 1052}

\bibitem[\protect\citeauthoryear{{Podsiadlowski}, {Joss}  \&
  {Hsu}}{{Podsiadlowski} et~al.}{1992}]{Podsiadlowski1992}
{Podsiadlowski} P.,  {Joss} P.~C.,   {Hsu} J.~J.~L.,  1992, \mn@doi [\apj]
  {10.1086/171341}, \href
  {https://ui.adsabs.harvard.edu/abs/1992ApJ...391..246P} {391, 246}

\bibitem[\protect\citeauthoryear{{Portegies Zwart}, {McMillan}  \&
  {Gieles}}{{Portegies Zwart} et~al.}{2010}]{PortegiesZwart2010}
{Portegies Zwart} S.~F.,  {McMillan} S. L.~W.,   {Gieles} M.,  2010, \mn@doi
  [\araa] {10.1146/annurev-astro-081309-130834}, \href
  {https://ui.adsabs.harvard.edu/abs/2010ARA&A..48..431P} {48, 431}

\bibitem[\protect\citeauthoryear{{Potter}, {Chitre}  \& {Tout}}{{Potter}
  et~al.}{2012}]{Potter2012}
{Potter} A.~T.,  {Chitre} S.~M.,   {Tout} C.~A.,  2012, \mn@doi [\mnras]
  {10.1111/j.1365-2966.2012.21409.x}, \href
  {https://ui.adsabs.harvard.edu/abs/2012MNRAS.424.2358P} {424, 2358}

\bibitem[\protect\citeauthoryear{Prinja, Barlow  \& Howarth}{Prinja
  et~al.}{1990}]{Prinja1990}
Prinja R.~K.,  Barlow M.,   Howarth I.~D.,  1990, \mn@doi [Astrophysical
  Journal] {10.1086/169224}, 361, 607

\bibitem[\protect\citeauthoryear{{Puchwein}, {Haardt}, {Haehnelt}  \&
  {Madau}}{{Puchwein} et~al.}{2019}]{Puchwein2019}
{Puchwein} E.,  {Haardt} F.,  {Haehnelt} M.~G.,   {Madau} P.,  2019, \mn@doi
  [\mnras] {10.1093/mnras/stz222}, \href
  {https://ui.adsabs.harvard.edu/abs/2019MNRAS.485...47P} {485, 47}

\bibitem[\protect\citeauthoryear{{Puls}, {Vink}  \& {Najarro}}{{Puls}
  et~al.}{2008}]{Puls2008}
{Puls} J.,  {Vink} J.~S.,   {Najarro} F.,  2008, \mn@doi [\aapr]
  {10.1007/s00159-008-0015-8}, \href
  {https://ui.adsabs.harvard.edu/abs/2008A&ARv..16..209P} {16, 209}

\bibitem[\protect\citeauthoryear{{Puls}, {Sundqvist}  \& {Markova}}{{Puls}
  et~al.}{2015}]{Puls2015}
{Puls} J.,  {Sundqvist} J.~O.,   {Markova} N.,  2015, in {Meynet} G.,  {Georgy}
  C.,  {Groh} J.,   {Stee} P.,  eds,  IAU Symposium Vol. 307, New Windows on
  Massive Stars. pp 25--36 (\mn@eprint {arXiv} {1409.3582}),
  \mn@doi{10.1017/S174392131400622X}

\bibitem[\protect\citeauthoryear{{Ramachandran}, {Hainich}, {Hamann},
  {Oskinova}, {Shenar}, {Sander}, {Todt}  \& {Gallagher}}{{Ramachandran}
  et~al.}{2018a}]{Ramachandran2018}
{Ramachandran} V.,  {Hainich} R.,  {Hamann} W.~R.,  {Oskinova} L.~M.,  {Shenar}
  T.,  {Sander} A.~A.~C.,  {Todt} H.,   {Gallagher} J.~S.,  2018a, \mn@doi
  [\aap] {10.1051/0004-6361/201731093}, \href
  {https://ui.adsabs.harvard.edu/abs/2018A&A...609A...7R} {609, A7}

\bibitem[\protect\citeauthoryear{{Ramachandran}, {Hamann}, {Hainich},
  {Oskinova}, {Shenar}, {Sander}, {Todt}  \& {Gallagher}}{{Ramachandran}
  et~al.}{2018b}]{Ramachandran2018b}
{Ramachandran} V.,  {Hamann} W.~R.,  {Hainich} R.,  {Oskinova} L.~M.,  {Shenar}
  T.,  {Sander} A.~A.~C.,  {Todt} H.,   {Gallagher} J.~S.,  2018b, \mn@doi
  [\aap] {10.1051/0004-6361/201832816}, \href
  {https://ui.adsabs.harvard.edu/abs/2018A&A...615A..40R} {615, A40}

\bibitem[\protect\citeauthoryear{{Ramachandran} et~al.,}{{Ramachandran}
  et~al.}{2019}]{Ramachandran2019}
{Ramachandran} V.,  et~al., 2019, \mn@doi [\aap] {10.1051/0004-6361/201935365},
  \href {https://ui.adsabs.harvard.edu/abs/2019A&A...625A.104R} {625, A104}

\bibitem[\protect\citeauthoryear{{Renzo} et~al.,}{{Renzo}
  et~al.}{2019}]{Renzo2019}
{Renzo} M.,  et~al., 2019, \mn@doi [\aap] {10.1051/0004-6361/201833297}, \href
  {https://ui.adsabs.harvard.edu/abs/2019A&A...624A..66R} {624, A66}

\bibitem[\protect\citeauthoryear{{Rey} \& {Starkenburg}}{{Rey} \&
  {Starkenburg}}{2022}]{Rey2022}
{Rey} M.~P.,  {Starkenburg} T.~K.,  2022, \mn@doi [\mnras]
  {10.1093/mnras/stab3709}, \href
  {https://ui.adsabs.harvard.edu/abs/2022MNRAS.510.4208R} {510, 4208}

\bibitem[\protect\citeauthoryear{{Rieder}, {Dobbs}, {Bending}, {Liow}  \&
  {Wurster}}{{Rieder} et~al.}{2022}]{Rieder2022}
{Rieder} S.,  {Dobbs} C.,  {Bending} T.,  {Liow} K.~Y.,   {Wurster} J.,  2022,
  \mn@doi [\mnras] {10.1093/mnras/stab3425}, \href
  {https://ui.adsabs.harvard.edu/abs/2022MNRAS.509.6155R} {509, 6155}

\bibitem[\protect\citeauthoryear{{Robertson}, {Ellis}, {Furlanetto}  \&
  {Dunlop}}{{Robertson} et~al.}{2015}]{Robertson2015}
{Robertson} B.~E.,  {Ellis} R.~S.,  {Furlanetto} S.~R.,   {Dunlop} J.~S.,
  2015, \mn@doi [\apjl] {10.1088/2041-8205/802/2/L19}, \href
  {https://ui.adsabs.harvard.edu/abs/2015ApJ...802L..19R} {802, L19}

\bibitem[\protect\citeauthoryear{{Romano}, {Karakas}, {Tosi}  \&
  {Matteucci}}{{Romano} et~al.}{2010}]{Romano10}
{Romano} D.,  {Karakas} A.~I.,  {Tosi} M.,   {Matteucci} F.,  2010, \mn@doi
  [\aap] {10.1051/0004-6361/201014483}, \href
  {https://ui.adsabs.harvard.edu/abs/2010A&A...522A..32R} {522, A32}

\bibitem[\protect\citeauthoryear{{Rosdahl}, {Schaye}, {Dubois}, {Kimm}  \&
  {Teyssier}}{{Rosdahl} et~al.}{2017}]{Rosdahl2017}
{Rosdahl} J.,  {Schaye} J.,  {Dubois} Y.,  {Kimm} T.,   {Teyssier} R.,  2017,
  \mn@doi [\mnras] {10.1093/mnras/stw3034}, \href
  {https://ui.adsabs.harvard.edu/abs/2017MNRAS.466...11R} {466, 11}

\bibitem[\protect\citeauthoryear{{Rosdahl} et~al.,}{{Rosdahl}
  et~al.}{2018}]{Rosdahl2018}
{Rosdahl} J.,  et~al., 2018, \mn@doi [\mnras] {10.1093/mnras/sty1655}, \href
  {https://ui.adsabs.harvard.edu/abs/2018MNRAS.479..994R} {479, 994}

\bibitem[\protect\citeauthoryear{Rosen, Lopez, Krumholz  \& Ramirez-Ruiz}{Rosen
  et~al.}{2014}]{Rosen2014}
Rosen A.~L.,  Lopez L.~A.,  Krumholz M.~R.,   Ramirez-Ruiz E.,  2014, \mn@doi
  [Monthly Notices of the Royal Astronomical Society] {10.1093/mnras/stu1037},
  442, 2701

\bibitem[\protect\citeauthoryear{{Sabhahit}, {Vink}, {Higgins}  \&
  {Sander}}{{Sabhahit} et~al.}{2021}]{Sabhahit2021}
{Sabhahit} G.~N.,  {Vink} J.~S.,  {Higgins} E.~R.,   {Sander} A. A.~C.,  2021,
  \mn@doi [\mnras] {10.1093/mnras/stab1948}, \href
  {https://ui.adsabs.harvard.edu/abs/2021MNRAS.506.4473S} {506, 4473}

\bibitem[\protect\citeauthoryear{{Sana} et~al.,}{{Sana}
  et~al.}{2012}]{Sana2012}
{Sana} H.,  et~al., 2012, \mn@doi [Science] {10.1126/science.1223344}, \href
  {https://ui.adsabs.harvard.edu/abs/2012Sci...337..444S} {337, 444}

\bibitem[\protect\citeauthoryear{{Sander} \& {Vink}}{{Sander} \&
  {Vink}}{2020}]{sander2020}
{Sander} A. A.~C.,  {Vink} J.~S.,  2020, \mn@doi [\mnras]
  {10.1093/mnras/staa2712}, \href
  {https://ui.adsabs.harvard.edu/abs/2020MNRAS.499..873S} {499, 873}

\bibitem[\protect\citeauthoryear{{Saviane}, {Rizzi}, {Held}, {Bresolin}  \&
  {Momany}}{{Saviane} et~al.}{2002}]{Saviane2002}
{Saviane} I.,  {Rizzi} L.,  {Held} E.~V.,  {Bresolin} F.,   {Momany} Y.,  2002,
  \mn@doi [\aap] {10.1051/0004-6361:20020750}, \href
  {https://ui.adsabs.harvard.edu/abs/2002A&A...390...59S} {390, 59}

\bibitem[\protect\citeauthoryear{{Schaerer}, {Fragos}  \& {Izotov}}{{Schaerer}
  et~al.}{2019}]{Schaerer2019}
{Schaerer} D.,  {Fragos} T.,   {Izotov} Y.~I.,  2019, \mn@doi [\aap]
  {10.1051/0004-6361/201935005}, \href
  {https://ui.adsabs.harvard.edu/abs/2019A&A...622L..10S} {622, L10}

\bibitem[\protect\citeauthoryear{{Schmutz}, {Leitherer}  \&
  {Gruenwald}}{{Schmutz} et~al.}{1992}]{Schmutz1992}
{Schmutz} W.,  {Leitherer} C.,   {Gruenwald} R.,  1992, \mn@doi [\pasp]
  {10.1086/133104}, \href
  {https://ui.adsabs.harvard.edu/abs/1992PASP..104.1164S} {104, 1164}

\bibitem[\protect\citeauthoryear{{Schootemeijer}, {Langer}, {Grin}  \&
  {Wang}}{{Schootemeijer} et~al.}{2019}]{Schootemeijer2019}
{Schootemeijer} A.,  {Langer} N.,  {Grin} N.~J.,   {Wang} C.,  2019, \mn@doi
  [\aap] {10.1051/0004-6361/201935046}, \href
  {https://ui.adsabs.harvard.edu/abs/2019A&A...625A.132S} {625, A132}

\bibitem[\protect\citeauthoryear{{Schootemeijer} et~al.,}{{Schootemeijer}
  et~al.}{2021}]{Schootemeijer2021}
{Schootemeijer} A.,  et~al., 2021, \mn@doi [\aap]
  {10.1051/0004-6361/202038789}, \href
  {https://ui.adsabs.harvard.edu/abs/2021A&A...646A.106S} {646, A106}

\bibitem[\protect\citeauthoryear{{Scott}, {Hirschi}, {Georgy}, {Arnett},
  {Meakin}, {Kaiser}, {Ekstr{\"o}m}  \& {Yusof}}{{Scott}
  et~al.}{2021}]{Scott2021}
{Scott} L.~J.~A.,  {Hirschi} R.,  {Georgy} C.,  {Arnett} W.~D.,  {Meakin} C.,
  {Kaiser} E.~A.,  {Ekstr{\"o}m} S.,   {Yusof} N.,  2021, \mn@doi [\mnras]
  {10.1093/mnras/stab752}, \href
  {https://ui.adsabs.harvard.edu/abs/2021MNRAS.503.4208S} {503, 4208}

\bibitem[\protect\citeauthoryear{{Searle}}{{Searle}}{1971}]{Searle1971}
{Searle} L.,  1971, \mn@doi [\apj] {10.1086/151090}, \href
  {https://ui.adsabs.harvard.edu/abs/1971ApJ...168..327S} {168, 327}

\bibitem[\protect\citeauthoryear{{Senchyna}, {Stark}, {Mirocha}, {Reines},
  {Charlot}, {Jones}  \& {Mulchaey}}{{Senchyna} et~al.}{2020}]{Senchyna2020}
{Senchyna} P.,  {Stark} D.~P.,  {Mirocha} J.,  {Reines} A.~E.,  {Charlot} S.,
  {Jones} T.,   {Mulchaey} J.~S.,  2020, \mn@doi [\mnras]
  {10.1093/mnras/staa586}, \href
  {https://ui.adsabs.harvard.edu/abs/2020MNRAS.494..941S} {494, 941}

\bibitem[\protect\citeauthoryear{{Shenar}, {Gilkis}, {Vink}, {Sana}  \&
  {Sander}}{{Shenar} et~al.}{2020}]{Shenar20}
{Shenar} T.,  {Gilkis} A.,  {Vink} J.~S.,  {Sana} H.,   {Sander} A.~A.~C.,
  2020, \mn@doi [\aap] {10.1051/0004-6361/201936948}, \href
  {https://ui.adsabs.harvard.edu/abs/2020A&A...634A..79S} {634, A79}

\bibitem[\protect\citeauthoryear{{Singh} et~al.,}{{Singh}
  et~al.}{2022}]{Singh2022}
{Singh} S.,  et~al., 2022, \mn@doi [Nature Astronomy]
  {10.1038/s41550-022-01610-5}, \href
  {https://ui.adsabs.harvard.edu/abs/2022NatAs...6..607S} {6, 607}

\bibitem[\protect\citeauthoryear{{Smartt}}{{Smartt}}{2009}]{Smartt2009}
{Smartt} S.~J.,  2009, \mn@doi [\araa] {10.1146/annurev-astro-082708-101737},
  \href {https://ui.adsabs.harvard.edu/abs/2009ARA&A..47...63S} {47, 63}

\bibitem[\protect\citeauthoryear{{Smith}, {Norris}  \& {Crowther}}{{Smith}
  et~al.}{2002}]{Smith2002}
{Smith} L.~J.,  {Norris} R. P.~F.,   {Crowther} P.~A.,  2002, \mn@doi [\mnras]
  {10.1046/j.1365-8711.2002.06042.x}, \href
  {https://ui.adsabs.harvard.edu/abs/2002MNRAS.337.1309S} {337, 1309}

\bibitem[\protect\citeauthoryear{{Smith}, {Bryan}, {Somerville}, {Hu},
  {Teyssier}, {Burkhart}  \& {Hernquist}}{{Smith} et~al.}{2021}]{Smith2021}
{Smith} M.~C.,  {Bryan} G.~L.,  {Somerville} R.~S.,  {Hu} C.-Y.,  {Teyssier}
  R.,  {Burkhart} B.,   {Hernquist} L.,  2021, \mn@doi [\mnras]
  {10.1093/mnras/stab1896}, \href
  {https://ui.adsabs.harvard.edu/abs/2021MNRAS.506.3882S} {506, 3882}

\bibitem[\protect\citeauthoryear{{Spruit}}{{Spruit}}{2002}]{Spruit2002}
{Spruit} H.~C.,  2002, \aap, 381, 923

\bibitem[\protect\citeauthoryear{{Steidel}, {Strom}, {Pettini}, {Rudie},
  {Reddy}  \& {Trainor}}{{Steidel} et~al.}{2016}]{Steidel2016}
{Steidel} C.~C.,  {Strom} A.~L.,  {Pettini} M.,  {Rudie} G.~C.,  {Reddy} N.~A.,
    {Trainor} R.~F.,  2016, \mn@doi [\apj] {10.3847/0004-637X/826/2/159}, \href
  {https://ui.adsabs.harvard.edu/abs/2016ApJ...826..159S} {826, 159}

\bibitem[\protect\citeauthoryear{{Sukhbold} \& {Adams}}{{Sukhbold} \&
  {Adams}}{2020}]{Sukhbold2020}
{Sukhbold} T.,  {Adams} S.,  2020, \mn@doi [\mnras] {10.1093/mnras/staa059},
  \href {https://ui.adsabs.harvard.edu/abs/2020MNRAS.492.2578S} {492, 2578}

\bibitem[\protect\citeauthoryear{Sutherland \& Dopita}{Sutherland \&
  Dopita}{1993}]{Sutherland1993}
Sutherland R.~S.,  Dopita M.~A.,  1993, \mn@doi [The Astrophysical Journal
  Supplement Series] {10.1086/191823}, 88, 253

\bibitem[\protect\citeauthoryear{{Sz{\'e}csi} \& {W{\"u}nsch}}{{Sz{\'e}csi} \&
  {W{\"u}nsch}}{2019}]{Szecsi2019}
{Sz{\'e}csi} D.,  {W{\"u}nsch} R.,  2019, \mn@doi [\apj]
  {10.3847/1538-4357/aaf4be}, \href
  {https://ui.adsabs.harvard.edu/abs/2019ApJ...871...20S} {871, 20}

\bibitem[\protect\citeauthoryear{{Sz{\'e}csi}, {Langer}, {Yoon}, {Sanyal}, {de
  Mink}, {Evans}  \& {Dermine}}{{Sz{\'e}csi} et~al.}{2015}]{Szecsi2015}
{Sz{\'e}csi} D.,  {Langer} N.,  {Yoon} S.-C.,  {Sanyal} D.,  {de Mink} S.,
  {Evans} C.~J.,   {Dermine} T.,  2015, \mn@doi [\aap]
  {10.1051/0004-6361/201526617}, \href
  {https://ui.adsabs.harvard.edu/abs/2015A&A...581A..15S} {581, A15}

\bibitem[\protect\citeauthoryear{{Sz{\'e}csi}, {Mackey}  \&
  {Langer}}{{Sz{\'e}csi} et~al.}{2018}]{Szecsi2018}
{Sz{\'e}csi} D.,  {Mackey} J.,   {Langer} N.,  2018, \mn@doi [\aap]
  {10.1051/0004-6361/201731500}, \href
  {https://ui.adsabs.harvard.edu/abs/2018A&A...612A..55S} {612, A55}

\bibitem[\protect\citeauthoryear{{Takahashi} \& {Langer}}{{Takahashi} \&
  {Langer}}{2021}]{takahashi2021}
{Takahashi} K.,  {Langer} N.,  2021, \mn@doi [\aap]
  {10.1051/0004-6361/202039253}, \href
  {https://ui.adsabs.harvard.edu/abs/2021A&A...646A..19T} {646, A19}

\bibitem[\protect\citeauthoryear{{Townsley}, {Feigelson}, {Montmerle}, {Broos},
  {Chu}  \& {Garmire}}{{Townsley} et~al.}{2003}]{townsley03}
{Townsley} L.~K.,  {Feigelson} E.~D.,  {Montmerle} T.,  {Broos} P.~S.,  {Chu}
  Y.-H.,   {Garmire} G.~P.,  2003, \mn@doi [\apj] {10.1086/376692}, \href
  {https://ui.adsabs.harvard.edu/abs/2003ApJ...593..874T} {593, 874}

\bibitem[\protect\citeauthoryear{{Trebitsch} et~al.,}{{Trebitsch}
  et~al.}{2021}]{Trebitsch2021}
{Trebitsch} M.,  et~al., 2021, \mn@doi [\aap] {10.1051/0004-6361/202037698},
  \href {https://ui.adsabs.harvard.edu/abs/2021A&A...653A.154T} {653, A154}

\bibitem[\protect\citeauthoryear{{Vartanyan}, {Laplace}, {Renzo},
  {G{\"o}tberg}, {Burrows}  \& {de Mink}}{{Vartanyan}
  et~al.}{2021}]{Vartanyan2021}
{Vartanyan} D.,  {Laplace} E.,  {Renzo} M.,  {G{\"o}tberg} Y.,  {Burrows} A.,
  {de Mink} S.~E.,  2021, \mn@doi [\apjl] {10.3847/2041-8213/ac0b42}, \href
  {https://ui.adsabs.harvard.edu/abs/2021ApJ...916L...5V} {916, L5}

\bibitem[\protect\citeauthoryear{{Verliat}, {Hennebelle}, {Gonz{\'a}lez}, {Lee}
   \& {Geen}}{{Verliat} et~al.}{2022}]{Verliat2022}
{Verliat} A.,  {Hennebelle} P.,  {Gonz{\'a}lez} M.,  {Lee} Y.-N.,   {Geen} S.,
  2022, \mn@doi [\aap] {10.1051/0004-6361/202141765}, \href
  {https://ui.adsabs.harvard.edu/abs/2022A&A...663A...6V} {663, A6}

\bibitem[\protect\citeauthoryear{{Vila-Costas} \& {Edmunds}}{{Vila-Costas} \&
  {Edmunds}}{1992}]{VilaCostas1992}
{Vila-Costas} M.~B.,  {Edmunds} M.~G.,  1992, \mn@doi [\mnras]
  {10.1093/mnras/259.1.121}, \href
  {https://ui.adsabs.harvard.edu/abs/1992MNRAS.259..121V} {259, 121}

\bibitem[\protect\citeauthoryear{{Vink}}{{Vink}}{2022}]{Vink2022}
{Vink} J.~S.,  2022, ARA\&A, \href
  {https://ui.adsabs.harvard.edu/abs/2021arXiv210908164V} {p. arXiv:2109.08164}

\bibitem[\protect\citeauthoryear{{Vink} \& {Gr{\"a}fener}}{{Vink} \&
  {Gr{\"a}fener}}{2012}]{Vink2012}
{Vink} J.~S.,  {Gr{\"a}fener} G.,  2012, \apjl, 751, L34

\bibitem[\protect\citeauthoryear{Vink, de Koter  \& Lamers}{Vink
  et~al.}{2001}]{Vink2001}
Vink J.~S.,  de Koter A.,   Lamers H. J. G. L.~M.,  2001, \mn@doi [Astronomy
  and Astrophysics] {10.1051/0004-6361:20010127}, 369, 574

\bibitem[\protect\citeauthoryear{{Vink}, {Brott}, {Gr{\"a}fener}, {Langer}, {de
  Koter}  \& {Lennon}}{{Vink} et~al.}{2010}]{Vink2010}
{Vink} J.~S.,  {Brott} I.,  {Gr{\"a}fener} G.,  {Langer} N.,  {de Koter} A.,
  {Lennon} D.~J.,  2010, \mn@doi [\aap] {10.1051/0004-6361/201014205}, \href
  {https://ui.adsabs.harvard.edu/abs/2010A&A...512L...7V} {512, L7}

\bibitem[\protect\citeauthoryear{{Vink}, {Higgins}, {Sander}  \&
  {Sabhahit}}{{Vink} et~al.}{2021}]{Vink2021}
{Vink} J.~S.,  {Higgins} E.~R.,  {Sander} A. A.~C.,   {Sabhahit} G.~N.,  2021,
  \mn@doi [\mnras] {10.1093/mnras/stab842}, \href
  {https://ui.adsabs.harvard.edu/abs/2021MNRAS.504..146V} {504, 146}

\bibitem[\protect\citeauthoryear{Weaver, McCray, Castor, Shapiro  \&
  Moore}{Weaver et~al.}{1977}]{Weaver1977}
Weaver R.,  McCray R.,  Castor J.,  Shapiro P.,   Moore R.,  1977, \mn@doi [The
  Astrophysical Journal] {10.1086/155692}, 218, 377

\bibitem[\protect\citeauthoryear{{White} \& {Long}}{{White} \&
  {Long}}{1991}]{White1991}
{White} R.~L.,  {Long} K.~S.,  1991, \mn@doi [\apj] {10.1086/170073}, \href
  {https://ui.adsabs.harvard.edu/abs/1991ApJ...373..543W} {373, 543}

\bibitem[\protect\citeauthoryear{{Woosley} \& {Heger}}{{Woosley} \&
  {Heger}}{2006}]{Woosley2006}
{Woosley} S.~E.,  {Heger} A.,  2006, \mn@doi [\apj] {10.1086/498500}, \href
  {https://ui.adsabs.harvard.edu/abs/2006ApJ...637..914W} {637, 914}

\bibitem[\protect\citeauthoryear{{Worseck}, {Prochaska}, {Hennawi}  \&
  {McQuinn}}{{Worseck} et~al.}{2016}]{Worseck2016}
{Worseck} G.,  {Prochaska} J.~X.,  {Hennawi} J.~F.,   {McQuinn} M.,  2016,
  \mn@doi [\apj] {10.3847/0004-637X/825/2/144}, \href
  {https://ui.adsabs.harvard.edu/abs/2016ApJ...825..144W} {825, 144}

\bibitem[\protect\citeauthoryear{{Yoon} \& {Langer}}{{Yoon} \&
  {Langer}}{2005a}]{Yoon2005}
{Yoon} S.~C.,  {Langer} N.,  2005a, \mn@doi [\aap]
  {10.1051/0004-6361:20042542}, \href
  {https://ui.adsabs.harvard.edu/abs/2005A&A...435..967Y} {435, 967}

\bibitem[\protect\citeauthoryear{{Yoon} \& {Langer}}{{Yoon} \&
  {Langer}}{2005b}]{YoonLanger2005}
{Yoon} S.~C.,  {Langer} N.,  2005b, \mn@doi [\aap]
  {10.1051/0004-6361:20054030}, \href
  {https://ui.adsabs.harvard.edu/abs/2005A&A...443..643Y} {443, 643}

\bibitem[\protect\citeauthoryear{{Yung}, {Somerville}, {Finkelstein},
  {Popping}, {Dav{\'e}}, {Venkatesan}, {Behroozi}  \& {Ferguson}}{{Yung}
  et~al.}{2020}]{Yung2020}
{Yung} L.~Y.~A.,  {Somerville} R.~S.,  {Finkelstein} S.~L.,  {Popping} G.,
  {Dav{\'e}} R.,  {Venkatesan} A.,  {Behroozi} P.,   {Ferguson} H.~C.,  2020,
  \mn@doi [\mnras] {10.1093/mnras/staa1800}, \href
  {https://ui.adsabs.harvard.edu/abs/2020MNRAS.496.4574Y} {496, 4574}

\bibitem[\protect\citeauthoryear{{Zapartas} et~al.,}{{Zapartas}
  et~al.}{2017}]{Zapartas2017}
{Zapartas} E.,  et~al., 2017, \mn@doi [\aap] {10.1051/0004-6361/201629685},
  \href {https://ui.adsabs.harvard.edu/abs/2017A&A...601A..29Z} {601, A29}

\bibitem[\protect\citeauthoryear{{da Silva}, {Fumagalli}  \& {Krumholz}}{{da
  Silva} et~al.}{2012}]{daSilva2012}
{da Silva} R.~L.,  {Fumagalli} M.,   {Krumholz} M.,  2012, \mn@doi [\apj]
  {10.1088/0004-637X/745/2/145}, \href
  {https://ui.adsabs.harvard.edu/abs/2012ApJ...745..145D} {745, 145}

\bibitem[\protect\citeauthoryear{{de Mink}, {Pols}, {Langer}  \& {Izzard}}{{de
  Mink} et~al.}{2009}]{deMink2009}
{de Mink} S.~E.,  {Pols} O.~R.,  {Langer} N.,   {Izzard} R.~G.,  2009, \mn@doi
  [\aap] {10.1051/0004-6361/200913205}, \href
  {https://ui.adsabs.harvard.edu/abs/2009A&A...507L...1D} {507, L1}

\bibitem[\protect\citeauthoryear{{ud-Doula} \& {Owocki}}{{ud-Doula} \&
  {Owocki}}{2002}]{ud-Doula2002}
{ud-Doula} A.,  {Owocki} S.~P.,  2002, \mn@doi [\apj] {10.1086/341543}, \href
  {https://ui.adsabs.harvard.edu/abs/2002ApJ...576..413U} {576, 413}

\bibitem[\protect\citeauthoryear{{van Zee} \& {Haynes}}{{van Zee} \&
  {Haynes}}{2006}]{vanZee2006}
{van Zee} L.,  {Haynes} M.~P.,  2006, \mn@doi [\apj] {10.1086/498017}, \href
  {https://ui.adsabs.harvard.edu/abs/2006ApJ...636..214V} {636, 214}

\makeatother
\end{thebibliography}

\end{document}